\documentclass[sigconf]{acmart}
\AtBeginDocument{%
  }

\setcopyright{acmlicensed}
\copyrightyear{2025}
\acmYear{2025}
\acmDOI{XXXXXXX.XXXXXXX}
\acmConference[Conference acronym 'XX]{Make sure to enter the correct
  conference title from your rights confirmation email}{June 03--05,
  2018}{Woodstock, NY}
\acmISBN{978-1-4503-XXXX-X/2018/06}

\usepackage[utf8]{inputenc} 
\usepackage[T1]{fontenc}    
\usepackage{hyperref}       
\usepackage{url}            
\usepackage{booktabs}       
\usepackage{amsfonts}       
\usepackage{nicefrac}       
\usepackage{microtype}      
\usepackage{xcolor} 

\usepackage{subfigure}

\usepackage{multirow}
\usepackage{tabularx}
\usepackage{adjustbox}
\usepackage{tikz}

\usepackage{wrapfig}
\usepackage{mathtools}
\usepackage[normalem]{ulem}
\usepackage{tabularx}
\usepackage{threeparttable}
\usepackage{amsmath,bm}
\usepackage{changepage}
\usepackage{mathtools}
\usepackage{makecell}
\usepackage{colortbl}

\begin{document}

\title{MIM: Multi-modal Content Interest Modeling Paradigm for User Behavior Modeling}


\author{
    Bencheng Yan$^{1*}$, Si Chen$^{1*}$, Shichang Jia$^{2}$, Jianyu Liu$^{1}$, 
    Yueran Liu$^{1}$, Chenghan Fu$^{1}$, Wanxian Guan$^{1}$, Hui Zhao$^{1}$, 
    Xiang Zhang$^{1}$, Kai Zhang$^{1}$, Wenbo Su$^{1}$, Pengjie Wang$^{1}$, 
    Jian Xu$^{1}$, Bo Zheng$^{1\dagger}$, Baolin Liu$^{2}$ \\
}

\thanks{$^{*}$ Bencheng Yan and Si Chen contributed equally to this work.}
\thanks{$^{\dagger}$ Bo Zheng(bozheng@alibaba-inc.com) is the corresponding author.}

\affiliation{%
  \institution{$^{1}$Alibaba Group, Beijing, China\\$^{2}$University of Science and Technology Beijing, Beijing, China}
  \city{}
  \state{}
  \country{}
}

\renewcommand{\shortauthors}{Yan et al.}

\begin{abstract}
  Click-Through Rate (CTR) prediction is a crucial task in recommendation systems, online searches, and advertising platforms, where accurately capturing users' real interests in content is essential for performance. However, existing methods heavily rely on ID embeddings, which fail to reflect users' true preferences for content such as images and titles. This limitation becomes particularly evident in cold-start and long-tail scenarios, where traditional approaches struggle to deliver effective results.
  To address these challenges, we propose a novel \textbf{M}ulti-modal Content \textbf{I}nterest \textbf{M}odeling paradigm (\textbf{MIM}), which consists of three key stages: Pre-training, Content-Interest-Aware Supervised Fine-Tuning (C-SFT), and Content-Interest-Aware UBM (CiUBM). 
  The pre-training stage adapts foundational models to domain-specific data, enabling the extraction of high-quality multi-modal embeddings. 
  The C-SFT stage bridges the semantic gap between content and user interests by leveraging user behavior signals to guide the alignment of embeddings with user preferences. 
  Finally, the CiUBM stage integrates multi-modal embeddings and ID-based collaborative filtering signals into a unified framework. Comprehensive offline experiments and online A/B tests conducted on the Taobao, one of the world's largest e-commerce platforms, demonstrated the effectiveness and efficiency of MIM method. The method has been successfully deployed online, achieving a significant increase of +14.14\% in CTR and +4.12\% in RPM, showcasing its industrial applicability and substantial impact on platform performance. To promote further research, we have publicly released the code and dataset at https://pan.quark.cn/s/8fc8ec3e74f3.  
\end{abstract}

\ccsdesc[500]{Information systems~Recommender systems}

\keywords{Multi-modal Representations, Click-Through Rate Prediction, Recommendation System, E-commerce Search}



\maketitle

\section{Introduction}
\label{sec:Introduction}
Click-Through Rate (CTR) prediction plays a vital role in applications such as recommendation systems, web searches, and online advertising\cite{covington2016Deep,yan2022apg}, as it directly impacts user engagement and platform revenue. Among its key components, User Behavior Modeling (UBM) has emerged as a critical optimization direction. By leveraging users' historical interactions, UBM effectively captures their underlying preferences, enabling more accurate predictions and enhancing overall system performance\cite{zhang2021deep,zhou2018deep,pi2020search}.

\begin{figure}[t]
\centering
\includegraphics[width = .4\textwidth]{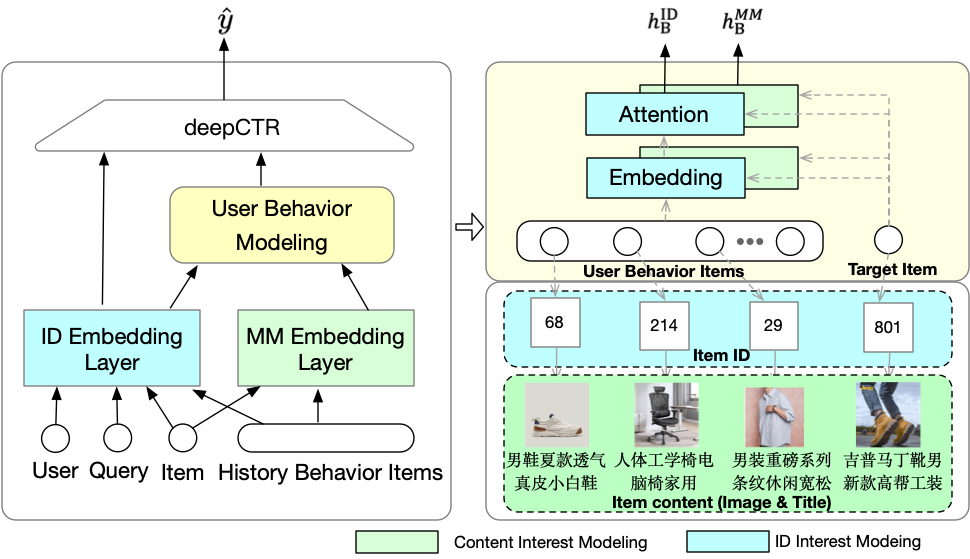}
\caption{An example of ID interest and content interest modeling in UBM.}
\label{figure:An example of ID interest and content interest modeling in UBM.}
\vspace{-2em}
\end{figure}

Traditional UBM methods predominantly rely on ID embeddings to represent items and user behaviors(see Figure\ref{figure:An example of ID interest and content interest modeling in UBM.} blue part). While effective in some scenarios, ID-based approaches face inherent limitations. First, ID embeddings primarily encode collaborative filtering (CF) signals but fail to effectively capture user preferences for content, such as images and titles, resulting in a misalignment between representations and actual user interests. Second, these methods require abundant user-item interaction data, resulting in poor performance in cold-start and long-tail scenarios\cite{yuan2021One,yuan2020ParameterEfficient}.

To overcome these limitations, there is a growing need for content-based multi-modal UBMs(see Figure\ref{figure:An example of ID interest and content interest modeling in UBM.} green part). This shift is motivated by two key factors. First, user interactions with recommendation systems are predominantly mediated through visual and textual content, such as product images and descriptions. Modeling these features is essential for accurately reflecting user preferences. Second, recent advancements in multi-modal foundation models (FoMs), such as Vision Transformers (ViT)\cite{dosovitskiy2020Image,liu2021Swin}, LLaMA\cite{touvron2023llamaopenefficientfoundation}, Vicuna\cite{vicuna2023}, BEiT-3\cite{wang2022image}, GPT-4\cite{openai2024gpt4technicalreport}, and so on, provide powerful tools for extracting rich semantic information from the visual and textual content.

Existing multi-modal user behavior modeling methods can be broadly categorized into two classes: two-stage pretraining methods and end-to-end training methods. Two-stage pretraining methods leverage pre-trained foundation models to extract multi-modal features, which are then integrated into user behavior models. End-to-end methods, on the other hand, jointly optimize multi-modal models and user behavior modeling modules, demonstrating their potential in better aligning user interests with content semantics. Despite achieving considerable success, these approaches also have limitations: two-stage methods lack deep alignment between multi-modal content and user interests, which constrains their effectiveness; end-to-end methods, while offering better alignment, come with high training costs, complex deployment requirements, and limited generalizability, making them less adaptable to diverse recommendation tasks. These limitations underscore the need for a universal and efficient approach to construct high-quality representations that align content with user interests.To address these challenges, three core questions must be answered. 

\noindent \textbf{$\bullet$ Q1: what kind of multi-modal embeddings are needed?} Pre-trained FoMs often fail to fully understand domain-specific content, and a semantic gap exists between content embeddings and user interests. 

\noindent \textbf{$\bullet$ Q2: how can multi-modal embeddings be effectively utilized in UBMs?} A unified framework is needed to integrate ID-based CF signals and multi-modal embeddings seamlessly. 

\noindent \textbf{$\bullet$ Q3: how can efficiency be ensured in large-scale applications?} Multi-modal embedding extraction often introduces high computational and memory costs, posing challenges for practical deployment.

To solve these challenges, this paper introduces MIM (Multi-modal Content Interest Modeling), a universal and efficient paradigm for multi-modal UBM. MIM adopts a three-stage framework to address the core questions comprehensively. In the pretraining stage, foundation models are adapted to downstream data to capture domain-specific content and align multi-modal features. In the content-interest-aware supervised fine-tuning (C-SFT) stage, user interest signals, such as purchase behaviors, are introduced to bridge the semantic gap and guide embeddings to reflect user preferences. In the CiUBM stage, a modular content-interest-aware UBM integrates ID-based CF signals, multi-modal content embeddings, and their interactions, providing an effective and flexible framework for user behavior modeling. To ensure efficiency, MIM introduces a representation center that precomputes and stores embeddings for fast retrieval, significantly reducing training and inference costs.

In summary, this paper makes the following contributions: (1) We propose a universal and effective multi-modal UBM method, MIM, which shifts the paradigm from ID-based to content-based interest modeling, enabling broad applicability and performance improvements across existing UBMs. (2) MIM demonstrates significant performance advantages, achieving lightweight training and efficient inference, making it scalable for industrial applications. (3) MIM has been successfully deployed in large-scale industrial scenarios, achieving 14.14\% CTR gains and 4.12\% RPM gains. To foster further research, we have publicly released the code and datasets.

\begin{figure*}[ht]
  \centering
  \includegraphics[width = 0.95\textwidth]{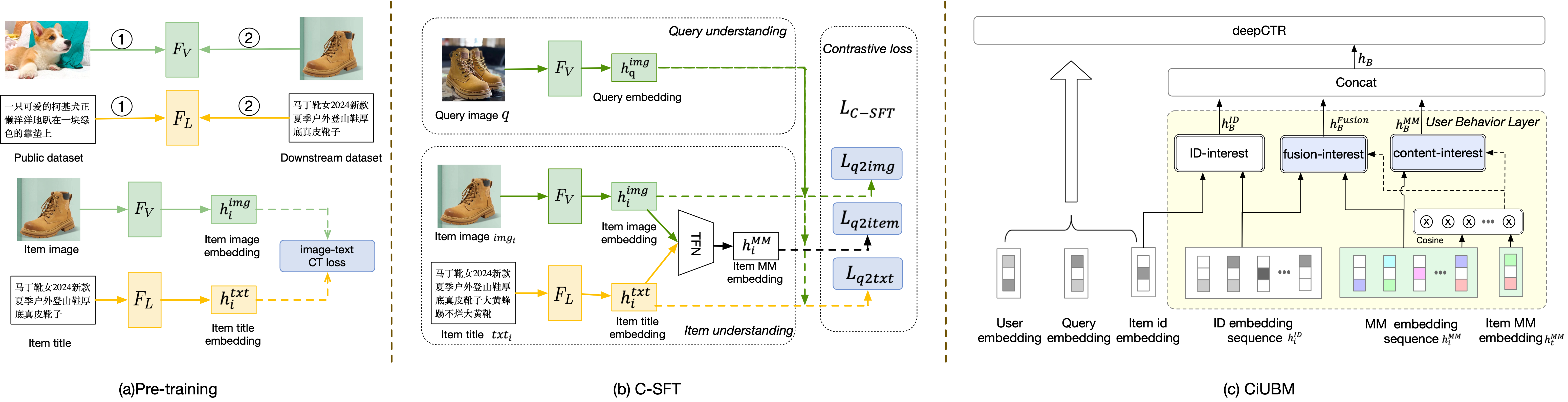}
  
  \caption{The framework of MIM. There are a total of three stages, including Pre-training, C-SFT, and CiUBM.
  Besides, a representation center is built for efficiency consideration.}
  
  \label{figure:The framework of MIM}
  \end{figure*}

\section{Related Works}
\label{sec:Related Work}
\subsection{User Behavior Modeling}
User Behavior Modeling (UBM) has been widely studied as a critical component of Click-Through Rate (CTR) prediction due to its ability to capture users' historical preferences and predict their future interactions. Traditional UBM methods, such as DIN\cite{zhou2018deep}, use attention mechanisms to selectively focus on relevant behaviors, while extensions like DIEN\cite{zhou2019deep} incorporate sequential modeling to capture the temporal evolution of user interests. Despite their success, these methods heavily rely on ID-based embeddings, which primarily encode collaborative filtering (CF) signals but fail to effectively model content-based preferences, particularly for items such as images and texts. This limitation becomes pronounced in cold-start and long-tail scenarios where interaction data is sparse\cite{yuan2021One,yuan2020ParameterEfficient}. Recent efforts, such as the introduction of hybrid UBM approaches, have attempted to integrate content features but still lack a unified framework for aligning content semantics with user-specific interests.
\subsection{Multi-modal Recommendation}
Multi-modal recommendation integrates diverse content features such as images, texts, and videos to improve user understanding and prediction accuracy. Traditional methods often augment CTR models with multi-modal (MM) features as additional attributes, directly feeding them into the models to enhance representation power \cite{mo2015image}. While effective to some extent, these approaches neglect the critical need to align MM features with user-specific interests, resulting in a limited understanding of user preferences.Recent advancements focus on bridging the domain gap between pre-trained foundation models and downstream recommendation tasks\cite{yuan2023Where, wang2023missrec, wu2021empowering, liu2023Multimodal}. However, these methods largely ignore the role of explicit user interest modeling in refining MM embeddings, limiting their ability to capture sophisticated user preferences.
Moreover, the practicality of existing methods in industrial applications remains a challenge. Many end-to-end frameworks incur high computational and memory costs, making them inefficient for large-scale deployment \cite{yuan2023Where}. 
To address these issues, our proposed MIM paradigm systematically bridges the semantic gap between MM features and user interests while maintaining scalability and efficiency, achieving significant performance gains in real-world applications.

\subsection{Foundation Models}
Pre-trained Foundation Models have significantly advanced the fields of Computer Vision (CV) and Natural Language Processing (NLP) by learning transferable representations from large-scale data. 
In CV, models like iGPT\cite{chen2020generative} and ViT\cite{dosovitskiy2020Image} have pioneered the application of transformer architectures for image recognition, leveraging self-supervised tasks such as masked patch prediction to capture rich visual semantics.
Further advancements, such as Swin Transformer\cite{liu2021Swin} and BEiT\cite{bao2021beit}, improved efficiency and scalability, making transformers a dominant paradigm in vision tasks.In NLP, foundational models like BERT\cite{devlin2019BERT},GPT-2\cite{radford2019language}, and XLNet\cite{yang2019xlnet} introduced pre-training on large corpora followed by task-specific fine-tuning, establishing a new standard for natural language understanding. These models have been extended to handle complex tasks, such as question answering, summarization, and sentiment analysis.
The integration of multi-modal learning has further broadened the applicability of foundation models. Methods like CLIP\cite{radford2021learning} propose contrastive pre-training techniques to align visual and textual modalities, enabling robust joint representations. 
Meanwhile, ViL-BERT\cite{lu2019vilbert} and LXMERT\cite{tan2019lxmert} extend the transformer framework to learn cross-modal interactions, achieving state-of-the-art performance on vision-language tasks. More recent models such as Flamingo\cite{alayrac2022flamingo} and BEiT-3\cite{wang2022image} demonstrate the effectiveness of scaling multi-modal architectures for a variety of downstream applications.

Despite these advancements, challenges remain in achieving efficient training and deployment of multi-modal foundation models, particularly in scenarios requiring alignment between diverse modalities and scalability in large-scale applications.

\section{MIM}

\subsection{Preliminaries and Method Overview}
\label{sec:Preliminaries and Method Overview}

\subsubsection{Preliminaries.}

\noindent \textbf{\newline CTR Prediction.} Considering a typical item search scenario\footnote{Note that in this section, we take the item search scenario as an example. Our proposed paradigm can also be applied in other industrial applications. Experiments all show the effectiveness of MIM on various scenarios \ref{sec:Evaluation on Industrial Applications}).}, given a set of users $\mathcal{U}$, a set of items $\mathcal{I}$, and a set of queries $\mathcal{Q}$, the CTR prediction task is to predict whether a user $u \in \mathcal{U}$ will click the item $i \in \mathcal{I}$ when $u$ searches a query $q \in \mathcal{Q}$.
It can be formulated as:
$\hat{y}=f(u,i,q)$
where $\hat{y}$ is the predicted click-through rate score.
Note that query $q$ can be a text query and can also be an image query, which refers to a user wanting to search for an item with similar content to the image query.

\noindent \textbf{User Behavior Modeling. }
Given user behavior $B=\{b_1,b_2,...,b_l\}$, where $l$ is the behavior length, a general UBM is expressed as 
\begin{align}
h_{B}^{ID} = \sum_{i=1}\alpha_{t,i}^{ID}h^{ID}_i
\label{eqn:UBM}
\end{align}
where $h^{ID}_i$ is the ID embedding of item $b_i$, $\alpha_{t,i}$ is the relevant score between $b_i$ and target item $t$ via attention
mechanism based on item ID embeddings.
Then $h^{ID}_B$ is fed into deepCTR.
In this way, the relative user interest to target item $t$ can be captured.

\subsubsection{Method Overview.}
The overall framework is illustrated in Figure \ref{figure:The framework of MIM}. The proposed MIM paradigm comprises three key stages designed to address the limitations of existing approaches and enhance user behavior modeling. In the pre-training stage, the primary focus is on adapting foundational models to domain-specific data, ensuring the extraction of high-quality multi-modal embeddings capable of understanding diverse item content. Subsequently, the Content-Interest-Aware Supervised Fine-Tuning (C-SFT) stage bridges the semantic gap between user interests and content representations by leveraging explicit user behavior signals, such as purchase actions, to guide embeddings in aligning with user preferences. Finally, the CiUBM stage integrates the refined multi-modal embeddings with ID-based collaborative filtering signals into a unified and flexible framework, enhancing the overall representation power. To address efficiency challenges in large-scale industrial applications, a representation center is introduced, which precomputes and stores embeddings for efficient retrieval, significantly reducing both training and inference costs.

\subsection{Pre-training}
\label{sec:Pre-train}
The pre-training stage focuses on equipping the multi-modal embeddings with the ability to understand item content effectively. This stage leverages foundational models (FoMs) pre-trained on public datasets (e.g., ImageNet and Wikipedia), which are further adapted to downstream domains. We define vision-based and language-based pre-trained FoMs as $F_V$ and $F_L$ respectively. To address the limitations of general pre-trained models, two aspects of adaptation need to be considered.: 

\noindent \textbf{Downstream Data Adaptation (DDA).} Knowledge learned from generic datasets may not align well with domain-specific item content. To overcome this, $F_V$ and $F_L$ are continually pre-trained on downstream datasets containing item images and textual descriptions. This process ensures that the resulting models are fine-tuned to the unique characteristics of domain-specific data, enabling a deeper understanding of visual and textual content.

\noindent \textbf{Different Modal Alignment (DMA).} Effective multi-modal representation requires a seamless alignment between embeddings from different modalities. Following the contrastive learning framework proposed in \cite{radford2021learning}, we perform image-text contrastive pre-training on downstream data. This process aligns image and text embeddings, ensuring a coherent representation that bridges the semantic gap between modalities.

By achieving these adaptation, the pre-training stage lays the foundation for robust multi-modal embeddings capable of effectively representing diverse item content.

\subsection{Content-Interest-Aware Supervised Fine-tuning} 
\label{sec:C-SFT}

After the pre-training stage, FoM can understand what content an item has.
However, To enhance its effectiveness in user behavior modeling, the focus of embeddings must shift from representing "what the content is" to "what users are interested in.
Ideally, multi-modal embeddings for items with similar user interests should be drawn closer together in the latent space, while embeddings for unrelated items should be pushed farther apart.
Inspired by the recent advancements in contrastive learning \cite{DBLP:journals/mta/LongDL24,DBLP:conf/cvpr/WuXYL18,DBLP:conf/iclr/ClarkLLM20,DBLP:conf/nips/MikolovSCCD13}, we propose a content-interest-aware supervised contrastive fine-tuning (C-SFT) method to effectively model user interest similarity within the multi-modal (MM) embedding space.

To design such a contrastive fine-tuning, we need to answer the following three questions:
\textbf{Q1}: How to define user interest pairs?
\textbf{Q2}: How to encode multi-modal features?
\textbf{Q3}: What is a proper learning objective?
Next, we present the design strategies for C-SFT by correspondingly answering the aforementioned questions.

\noindent \textbf{Q1: User Interest Pairs Definition. }

The success of the contrastive learning framework largely relies on the definition of the user interest pairs\cite{DBLP:conf/kdd/QiuCDZYDWT20}.
It drives us to seek a strong and direct user interest signal.

In this paper, we leveraged data from visual search scenario, we define the user interest pair as <an image query $q$, an item $i$>, which refers to a user who has searched an image query $q$ and purchased an item $i$. Compared to textual queries, image queries carry richer semantic meanings and can more accurately express users' intentions and interests

The reason to use purchase behaviors rather than click behaviors is that purchase actions provide a clearer and more reliable indication of user interests.
We also conduct experiments to show the impacts of different signals in Section \ref{sec:Effectiveness of C-SFT}.

\noindent \textbf{Q2: Multi-modal Encoder. }
Given the user interest pairs <$q$, $i$>, a multi-modal encoder extracts embeddings for both the query and the item. The query image $q$ and item image $img_i$ are processed using $F_V$, while the item title $txt_i$ is encoded by $F_L$:
\begin{equation}
h^{img}_q = F_V(q), \quad h^{img}_i = F_V(img_i), \quad h^{txt}_i = F_L(txt_i)
\end{equation}

To integrate the item's multi-modal content, we fuse the image and title embeddings using a tensor-based approach inspired by TFN \cite{zadeh2017tensor}, capturing both intra- and inter-modal interactions. The fused embedding is defined as:
\begin{equation}
\hat{h}^{MM}_{i} = \begin{bmatrix} h^{img}_i \\ 1 \end{bmatrix} \otimes \begin{bmatrix} h^{txt}_i \\ 1 \end{bmatrix}
\end{equation}

where $\otimes$ indicates the outer product. Finally, the fused representation is refined using an MLP:
\begin{equation}
h^{MM}_i = MLP(\hat{h}^{MM}_{i})
\end{equation}
producing a comprehensive multi-modal embedding $h^{MM}_i$ that effectively represents the item's content.

\noindent \textbf{Q3: Loss Function Design. }
To model the similarity between the pair <$q$,$i$>, we adopt the standard InfoNCE loss function:
\begin{equation}
\label{equ:contrastive loss}
L_{q2item} = - \log \left( \frac{\exp(sim(h^{img}_q, (h^{MM}_i)^+))}{\sum_{j=0}^{K}\exp(sim(h^{img}_q, (h^{MM}_j)^-))} \right)
\end{equation}
where \( sim(h^{img}_q, (h^{MM}_i)^+) = \frac{(h^{img}_q)^\top (h^{MM}_i)^+}{\|h^{img}_q\| \| (h^{MM}_i)^+ \|} \), and \( K \) is the number of negative samples taken from other pairs in the same batch.

\begin{figure}[t]
\centering
\includegraphics[width = .45\textwidth]{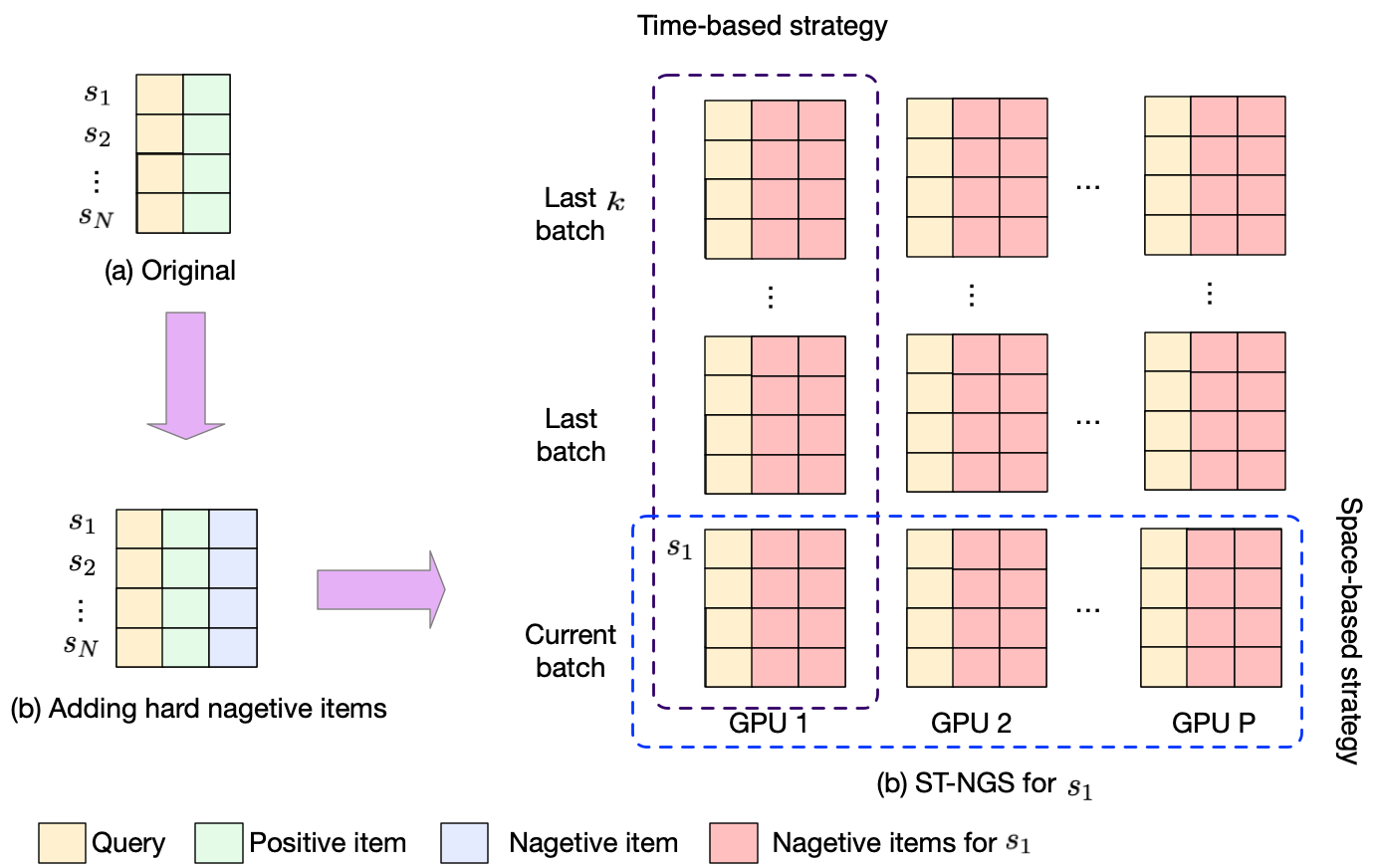}

\caption{The example of negative sample generation. (a) Negative items are obtained from other items in the same batch. (b) adding a hard negative item (b) Taking the sample $s_1$ as an example, with the help of ST-NGS, the amount of negative items (i.e., the red items) can be added.}

\label{figure:The example of negative sample generation.}
\end{figure}

However, there are two shortcomings in directly using the basic InfoNCE.: (1) The number of negative samples is limited by the batch size, which is insufficient for industrial-scale applications where billions of items exist. (2) Single-modal embeddings, such as \( h^{img}_i \) and \( h^{txt}_i \), are ignored, which may result in a loss of modality-specific details.To address these, we propose two extensions:

\noindent \textbf{Extension 1: Space-time-based negative sample generation (ST-NSG).}
As shown in Figure \ref{figure:The example of negative sample generation.}, to address the limitation of limited negative samples, we propose the Space-Time Negative Sampling (ST-NSG) method. First, for each sample $<q, i>$, we add a hard negative item $i^-$ from the same category, forming the triplet $<q, i, i^->$. To expand the negative sample set, we use two strategies: (1) Time-based strategy: Negative items are collected from the current batch and the last $k$ batches, increasing the number of negatives to $2N(k+1)-1$, where $N$ is the batch size. (2) Space-based strategy: In distributed training, items from $P$ GPUs across $k$ recent batches are added as negatives, resulting in $2NP(k+1)-1$ total negatives. Combined, these strategies provide nearly $200\times$ more negative samples compared to basic InfoNCE, significantly enhancing model performance in practice.

\noindent \textbf{Extension 2: Multi-level InfoNCE.}
Although MM embeddings provide a holistic view of item content across modalities, they may lack details specific to each modality. To address this, we propose \textbf{multi-level InfoNCE}, which incorporates signals from both single-modality and multi-modal embeddings. The overall loss function is defined as:

\begin{equation}
\label{equ:sft loss}
L_{C-SFT} = L_{q2item} + \alpha L_{q2txt} + \beta L_{q2img}
\end{equation}

where $L_{q2txt}$ and $L_{q2img}$ represent the InfoNCE loss for the pairs $<h^{img}_q, h^{txt}_i>$ and $<h^{img}_q, h^{img}_i>$, respectively. The weights $\alpha$ and $\beta$ balance the contributions of each modality, ensuring a more comprehensive modeling of user interests.

\subsection{CiUBM: Content-interest-aware UBM}
\label{Sec:CiUBM}
After obtaining the multi-modal (MM) embedding $h^{MM}_i$ for item $i$, we design a content-interest-aware user behavior model (CiUBM) to enhance the existing UBMs by integrating ID and content-based signals. CiUBM consists of three components:

\noindent \textbf{ID Interest Module. } This module uses existing UBM techniques to model user preferences based on ID embeddings, capturing collaborative filtering (CF) signals. It provides a straightforward integration of prior UBM methods.

\noindent \textbf{Content Interest Module. } This module leverages the high-quality MM embeddings to calculate user content interest relative to the target item $t$. Similarity scores $\alpha_{t,i}^{MM}$ between the target and historical behavior items are used to weight the MM embeddings, creating a representation $h_{B}^{MM}$ that reflects user content preferences. Formally:
\begin{equation}
\alpha_{t,i}^{MM} = \frac{(h^{MM}_t)^\top h^{MM}_i}{\|h^{MM}_t\| \|h^{MM}_i\|}, \quad h_{B}^{MM} = \sum_{i=1}^l \alpha_{t,i}^{MM} h^{MM}_i
\end{equation}
where $h^{MM}_i$ is the MM embedding of the historical item $b_i$, and $l$ is the number of user behaviors.

\noindent \textbf{Fusion Interest Module. } This module combines ID and content interest. By using $\alpha_{t,i}^{MM}$ to weight ID embeddings $h_{i}^{ID}$, it produces $h_{B}^{Fusion}$, which aligns content-based and collaborative filtering signals:
\begin{equation}
h_{B}^{Fusion} = \sum_{i=1}^l \alpha_{t,i}^{MM} h_{i}^{ID}
\end{equation}

\noindent \textbf{Final Representation. } The overall user behavior representation $h_B$ is obtained by concatenating the outputs of the three modules:
\begin{equation}
h_B = \textit{Concat}(h_{B}^{ID}, h_{B}^{MM}, h_{B}^{Fusion})
\end{equation}
and is then fed into the deepCTR model for improved prediction.

\subsection{Representation Center}
\label{sec:Representation Center}
\begin{figure}[t]
\centering
\includegraphics[trim=10 0 0 0, clip,width = .45\textwidth]{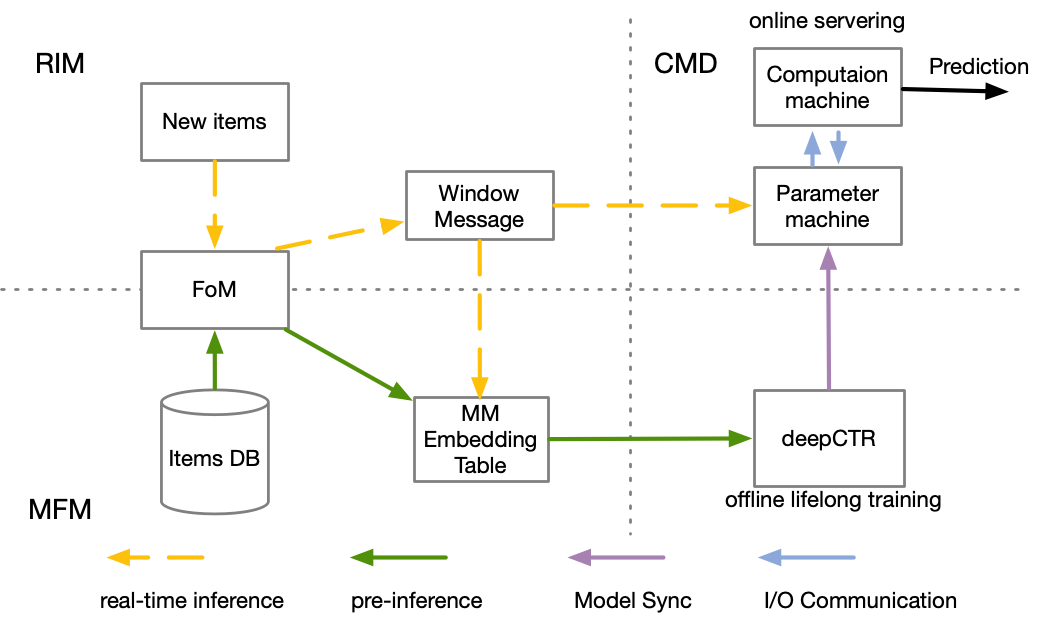}
\vspace{-1em}
\caption{The framework of representation center.}
\vspace{-1em}
\label{figure:The framework of representation_center.}
\end{figure}

In general, the decomposed paradigm allows MIM to train different stage models in parallel to save time cost.
However, the MM embedding extraction in CiUBM stage will definitely consume more time and GPU memory.
Even worse, all items in user behavior $B$ are needed to be extracted, aggravating the time and GPU memory cost (see Section \ref{sec:Efficiency Evaluation}).
To address this problem, we proposed the representation center, which contains three key components (Figure \ref{figure:The framework of representation_center.}).

\noindent \textbf{MM Features Memorization (MFM). }
Rather than directly inferring the MM embedding during deepCTR training or inferencing, we pre-infer the MM embeddings of all items.
Then we memorize these features as an embedding table, where the key is the item ID and the value is the MM embedding.
In this way, we can directly obtain MM embedding from the embedding table without any extra time and GPU memory cost.

\noindent \textbf{Real-time Inference Module (RIM). }
The key problem of maintaining the MM embedding table is that it cannot obtain the MM features of new items in time.
However, in practice, it is a very common situation where new items are shown all the time.
Thus, RIM is designed to address this problem.
Specifically, when a new item is added to industrial applications, RIM will infer the MM embedding of this new item in real-time.
Then, this inferred MM embedding will be collected as a window message, which is sent to the MM embedding table for offline training and to the parameter machine (see the following part for details) for online serving in a short window time.

\noindent \textbf{Computation and Memory Decomposition (CMD). }
Although introducing MFM will not cost too much GPU memory, the embedding table still needs a large amount of random access memory (e.g., nearly 300G in our experiments), which also influences the efficiency (e.g., the computation of some operations in CPU and I/O between GPU and CPU).
To solve this problem, we propose a separate strategy where the model parameters (including embedding tables and dense layer parameters) and the computation of inference are divided into different machines, i.e., the parameter machine and the computation machine.
In this way, the computation machine only needs to care about the memory cost during the feed-forward and is not restricted by the huge embedding tables\footnote{During inference, the computation machine can only request the embedding vectors of the corresponding keys rather than the whole embedding table to save memory}.
Meanwhile, with the help of MFM, there are no additional parameters needed for loading and maintaining a number of temporary tensors of MM feature extraction by large-scale FoM.

\section{Experiments}

\subsection{Experimental Settings}

\subsubsection{Datasets.}
\emph{(1) Pre-training Datasets.} 
As introduced in Section \ref{sec:Pre-train}, a downstream-related dataset is applied to align the style and modal distribution.
We collect images and titles of items from Taobao, one of the world's largest online retail platforms.
There are a total of 1.42+B items.
\emph{(2) C-SFT Datasets.}
We collect purchase logs in the last 6 months from the visual search scenario in Taobao and construct the C-SFT dataset as introduced in Section \ref{sec:C-SFT}.
There are a total of 0.14 B user interest pairs.
\emph{(3) Downstream Datasets.}
The dataset is collected from the real-world traffic logs in Taobao.
Each sample refers to a user who searches a query on this platform and observes an item and contains the necessary information about the queries and items, including multi-modal embedding, product IDs, categories, brands, prices, sales, etc.
The label is defined as whether the user clicks this item.
Besides, we further adopt 7 subsets to sufficiently evaluate the performance of MIM.
More detailed information about the dataset is summarized in Table \ref{tab:The statistic of datasets.}.
Note that, due to the limited space, apart from Section \ref{sec:Performance on CTR Prediction Tasks}, we report the results on the whole downstream dataset by default, and similar conclusions can be obtained from other sub-datasets.

\begin{table}
\caption{The statistic of datasets.
B refers to billion, and M refers to million.}
\label{tab:The statistic of datasets.}
\centering
\resizebox{\columnwidth}{!}{
\begin{tabular}{c|cccccccc}
\toprule
                                                     & \multicolumn{1}{c|}{\textbf{ALL}} & \multicolumn{1}{c|}{\textbf{Home  Decoration}} & \multicolumn{1}{c|}{\textbf{Clothes}} & \multicolumn{1}{c|}{\textbf{Sports}} & \multicolumn{1}{c|}{\textbf{Toys}} & \multicolumn{1}{c|}{\textbf{FMCG}} & \multicolumn{1}{c|}{\textbf{Pets}} & \textbf{Cars} \\ \midrule
\#Data                                              & \multicolumn{1}{c|}{1.83+ B}    & \multicolumn{1}{c|}{0.15+ B}                  & \multicolumn{1}{c|}{0.91+ B}        & \multicolumn{1}{c|}{0.1+ B}       & \multicolumn{1}{c|}{51.6+ M}     & \multicolumn{1}{c|}{0.21+ B}               & \multicolumn{1}{c|}{24.81+ M}     &      60,75+ M
\\ 
\#Query                                              & \multicolumn{1}{c|}{56.34+ M}    & \multicolumn{1}{c|}{5.32+ M}                  & \multicolumn{1}{c|}{15.07+ M}        & \multicolumn{1}{c|}{3.56+ M}       & \multicolumn{1}{c|}{2.32+ M}     & \multicolumn{1}{c|}{6.72+ M}               & \multicolumn{1}{c|}{0.73+ M}     &      3,66+ M
\\ 
\#Users                                             & \multicolumn{1}{c|}{0.12+ B}    & \multicolumn{1}{c|}{14.04+ M}                  & \multicolumn{1}{c|}{55.77+ M}        & \multicolumn{1}{c|}{16.34+ M}       & \multicolumn{1}{c|}{7.04+ M}     & \multicolumn{1}{c|}{25.95+ M}               & \multicolumn{1}{c|}{2.63+ M}     &      6.62+ M
\\ 
\#Items            &                                     \multicolumn{1}{c|}{16.16+ M}    & \multicolumn{1}{c|}{1.50+ M}                  & \multicolumn{1}{c|}{3.74+ M}        & \multicolumn{1}{c|}{0.65+ M}       & \multicolumn{1}{c|}{0.42+ M}     & \multicolumn{1}{c|}{1.11+ M}               & \multicolumn{1}{c|}{0.22+ M}     &      0.91+ M\\ \bottomrule
\end{tabular}
}
\end{table}


\subsubsection{Baselines} Here, we compare our method with various UBM methods, including Avg Pooling\footnote{averaging the embedding of different items in user behaviors}, DIN\cite{zhou2018deep}, DIEN\cite{zhou2019deep}, BST\cite{chen2019behavior}, and SIM\cite{pi2020search}, TWIN\cite{chang2023twin}.
By default, the base backbone of UBM is set as SIM \cite{pi2020search}.

\subsubsection{Training Details.}
\label{sec:Training Details.}
We use the Adam optimizer with a learning rate of 0.005 for all methods.
The batch size N is 1024 for all methods.

By default, we take EVA-2 \cite{woo2023convnext} for images and BGE\cite{bge_embedding} for texts.
We set hyper-parameters $\alpha=0.5$, $\beta=0.5$ for loss weights and $k=10$ for negative samples by grid searching.
Besides, we also evaluate the impact of different FoM, including ResNet50\cite{he2019bag}, SwinTransformer\cite{liu2021Swin}, ConvNext v1\cite{liu2022ConvNeta}, ConvNext V2\cite{woo2023convnext}, ConvNext v2 large\cite{woo2023convnext}, EVA-2\cite{fang2023eva02}, Bert\cite{devlin2018bert}, GPT-2\cite{radford2019language}, BEiT-3\cite{wang2022image} and BGE\cite{bge_embedding} in Section\ref{sec:Effectiveness Evaluation}.

MIM is trained by 64 A100 GPUs in the pre-training stage, 40 A100 GPUs in the C-SFT stage, and 100 V100 GPUs in the CiUBM training stage.
We run all experiments multiple times with different random seeds and report the average results.

\begin{table*}[t] \footnotesize
\caption{The AUC results of the CTR prediction task.
Note Base refers to the original results of the corresponding methods and Base+MIM refers to the results
with the help of MIM.
$\Delta$ refers to the improvement of Base+MIM compared to Base.
}
\vspace{-1em} 
\label{tab:Performance on CTR Prediction Tasks}
\resizebox{0.9\textwidth}{!}{%
\centering
\begin{tabular}{l|l|c|c|c|c|c|c|c|c}
\toprule
\multicolumn{1}{l|}{\textbf{Model}}         & \multicolumn{1}{l|}{\textbf{Version}} & \textbf{ALL}             & \textbf{Toys}            & \textbf{FMCG}            & \textbf{Sports}          & \textbf{Cars}            & \textbf{Clothes}         & \textbf{Pets}            & \textbf{Home Decoration} \\ \midrule
\multirow{3}{*}{Avg Pooling} & Base                 & 0.6965          & 0.6958          & 0.7044          & 0.7045          & 0.6965          & 0.6810          & 0.6844          & 0.6748          \\
                             & Base+MIM             & 0.7030          & 0.6989          & 0.7168          & 0.7148          & 0.7065          & 0.6828          & 0.6911          & 0.6796          \\
                             & $\Delta$                & \textbf{+0.0066} & \textbf{+0.0032} & \textbf{+0.0123} & \textbf{+0.0104} & \textbf{+0.0100} & \textbf{+0.0017} & \textbf{+0.0067} & \textbf{+0.0048} \\ \midrule
\multirow{3}{*}{DIN}         & Base                 & 0.7004          & 0.6969          & 0.7187          & 0.6986          & 0.6928          & 0.6753          & 0.7099          & 0.6802          \\
                             & Base+MIM             & 0.7047          & 0.7006          & 0.7225          & 0.7027          & 0.6963          & 0.6812          & 0.7126          & 0.6864          \\
                             & $\Delta$                & \textbf{+0.0042} & \textbf{+0.0037} & \textbf{+0.0039} & \textbf{+0.0041} & \textbf{+0.0036} & \textbf{+0.0058} & \textbf{+0.0028} & \textbf{+0.0062} \\ \midrule
\multirow{3}{*}{DIEN}        & Base                 & 0.7010          & 0.6975          & 0.7190          & 0.6989          & 0.6934          & 0.6761          & 0.7102          & 0.6809          \\
                             & Base+MIM             & 0.7055          & 0.7013          & 0.7233          & 0.7033          & 0.6970          & 0.6822          & 0.7131          & 0.6872          \\
                             & $\Delta$                & \textbf{+0.0046} & \textbf{+0.0038} & \textbf{+0.0044} & \textbf{+0.0045} & \textbf{+0.0036} & \textbf{+0.0061} & \textbf{+0.0029} & \textbf{+0.0063} \\ \midrule
\multirow{3}{*}{BST}         & Base                 & 0.6995          & 0.6965          & 0.7177          & 0.6973          & 0.6924          & 0.6740          & 0.7093          & 0.6796          \\
                             & Base+MIM             & 0.7049          & 0.7000          & 0.7229          & 0.7026          & 0.6964          & 0.6807          & 0.7129          & 0.6861          \\
                             & $\Delta$                & \textbf{+0.0054} & \textbf{+0.0035} & \textbf{+0.0053} & \textbf{+0.0053} & \textbf{+0.0040} & \textbf{+0.0068} & \textbf{+0.0036} & \textbf{+0.0065} \\ \midrule
\multirow{3}{*}{TWIN}        & Base                 & 0.7028          & 0.6992          & 0.7214          & 0.7013          & 0.6951          & 0.6776          & 0.7126          & 0.6825          \\
                             & Base+MIM             & 0.7051          & 0.7010          & 0.7229          & 0.7032          & 0.6968          & 0.6813          & 0.7134          & 0.6867          \\
                             & $\Delta$                & \textbf{+0.0023} & \textbf{+0.0018} & \textbf{+0.0016} & \textbf{+0.0019} & \textbf{+0.0017} & \textbf{+0.0037} & \textbf{+0.0009} & \textbf{+0.0042} \\ \midrule
\multirow{3}{*}{SIM}         & Base                 & 0.7024          & 0.7017          & 0.7111          & 0.7024          & 0.6945          & 0.6869          & 0.7058          & 0.6928          \\
                             & Base+MIM             & 0.7062          & 0.7067          & 0.7153          & 0.7065          & 0.6993          & 0.6893          & 0.7112          & 0.6951          \\
                             & $\Delta$                & \textbf{+0.0038} & \textbf{+0.0050} & \textbf{+0.0042} & \textbf{+0.0041} & \textbf{+0.0048} & \textbf{+0.0024} & \textbf{+0.0054} & \textbf{+0.0023} \\
\bottomrule
\end{tabular}
}
\end{table*}

\subsection{Performance on CTR Prediction Tasks}
\label{sec:Performance on CTR Prediction Tasks}
We evaluated the performance of MIM on CTR prediction tasks by applying it to various existing UBM methods. As a universal paradigm, MIM is compatible with most UBM frameworks. To assess its effectiveness, we compared the AUC scores of the original models (denoted as Base) with those enhanced by MIM. The AUC metric \cite{fawcett2006introduction} is reported.
As shown in Table \ref{tab:Performance on CTR Prediction Tasks}, integrating MIM leads to consistent performance improvements across all baseline methods and datasets. For example, SIM achieves AUC gains of 0.23pt to 0.54pt, and similar gains are observed for other UBMs. While these improvements may appear modest, in large-scale scenarios such as Taobao, a 0.1pt improvement in AUC can result in several percentage points of uplift in online CTR. This uplift can bring billions in revenue to the platform, highlighting the economic value of our approach.
These results validate the effectiveness and generalizability of MIM. Which can significantly enhances user behavior modeling by bridging the gap between content and user interest embeddings. Furthermore, its universal design ensures compatibility with a wide range of existing UBM frameworks, making it a practical solution for industrial-scale CTR prediction tasks.
\vspace{-1em}

\subsection{Effectiveness Evaluation}
\label{sec:Effectiveness Evaluation}
In this section, we conduct various experiments to detailedly evaluate the effectiveness of MIM.

\subsubsection{Effectiveness of high-quality MM embeddings.}
\label{sec:Effectiveness of high-quality MM embeddings.}

\begin{figure}[t]

  \centering
      \subfigure[The impact of different $F_V$]{
     \includegraphics[trim=5 5 5 5, clip,width = .22\textwidth]{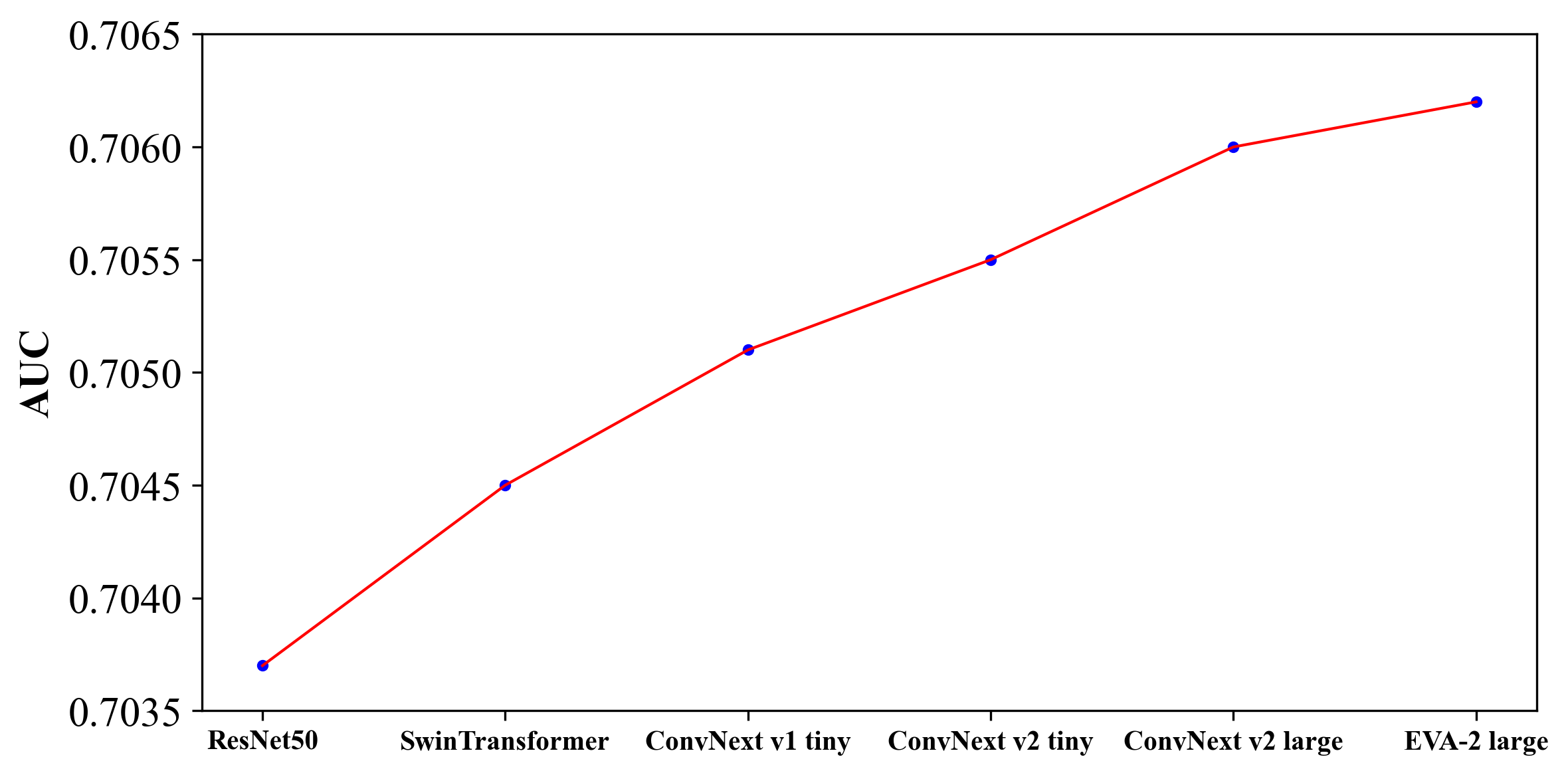}
     \label{fig:subfigure2}
   }
        \subfigure[The impact of different $F_L$]{
     \includegraphics[trim=5 5 5 5, clip,width = .22\textwidth]{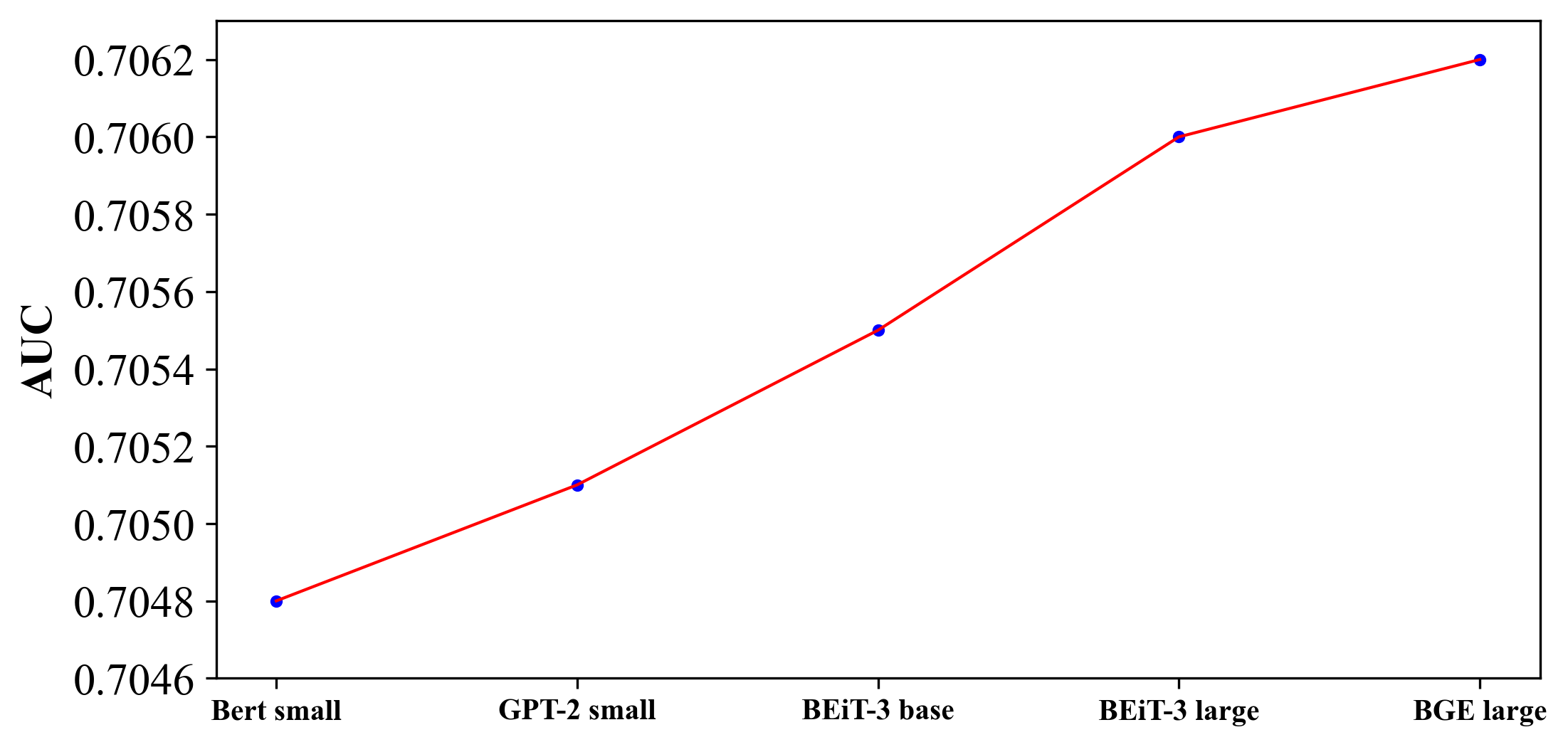}
     \label{fig:subfigure2}
   }
 \vspace{-1em}
 \caption{Evaluation of the impact of different FoMs including $F_V$ and $F_L$.
 From left to right, a more powerful FoM is adopted.}
 \vspace{-1em}
 \label{figure:Evaluation of the impact of different FoMs}
 \end{figure}

\noindent \textbf{\newline The impact of FoM. }
The experimental results in Figure \ref{figure:Evaluation of the impact of different FoMs} demonstrate the effectiveness of upgrading FoM. For $F_V$, the shift from ResNet50 to EVA-2 yields constant performance improvements, demonstrating the ability of stronger visual FoMs to capture richer content semantics. Similarly, for $F_L$, upgrading from smaller to larger models (e.g., BEiT-3 base to large) improves AUC, highlighting the benefits of scaling parameter sizes. These findings confirm that enhancing FoM, both in terms of model strength and parameter size, significantly improves the representation and alignment of user interests across modalities.

\begin{table}[t] \footnotesize
  \centering
  \caption{The impact of pre-training.
  }
  \vspace{-1em}
  \label{tab:The impact of pre-training}
  \resizebox{0.8\columnwidth}{!}{%
  \centering
  \begin{tabular}{l|c|c}
  \toprule
  \textbf{Version}                          & \textbf{AUC}    & \textbf{$\Delta$} \\ \midrule
  Base                      & 0.7024 &  - \\ \midrule
  Plain FoM                & 0.7030 & +0.0060          \\
  +DDA                      & 0.7034 & +0.0010    \\
  +DDA+DMA                  & 0.7040 & +0.0016    \\
  +DDA+DMA+C-SFT(i.e., MIM) & 0.7062 & +0.0038   \\
  \bottomrule
  \end{tabular}
  }
  \end{table}

\noindent \textbf{The impact of pre-training. }
We evaluate the impact of different strategies used in the pre-training stage.
The results are presented in Table \ref{tab:The impact of pre-training}.
Here, Plain FoM refers to the FoM that is only pre-trained on general datasets.
From Table \ref{tab:The impact of pre-training}, it shows (1) directly applying Plain FoM only achieves limited improvement due to the sub-effectiveness in item content understanding.
(2) Both downstream data adaption and different modal alignment can help FoM better capture the item content and provide a significant improvement.

\begin{table}[t]
  \footnotesize 
  \caption{The results of C-SFT evaluation. LF, CE, and CL refer to loss function, cross-entropy loss, and contrastive loss respectively.}
  \vspace{-1em}
  \centering
  \resizebox{0.8\columnwidth}{!}{%
  \label{tab:Effectiveness Evaluation}
  \begin{tabular}{l|l|l|c|c}
  \toprule
  \textbf{Version}  & \textbf{LF} & \textbf{Signal} & \textbf{AUC} & \textbf{$\Delta$} \\ \midrule
  Base              & -           & -               & 0.7024       & -                  \\ \midrule
  v1                & CE          & Category        & 0.7043       & +0.0019            \\
  v2                & CE          & Click           & 0.7053       & +0.0029            \\
  v3                & CE          & Purchase        & 0.7059       & +0.0035            \\
  v4                & CL          & Click           & 0.7055       & +0.0031            \\
  v5                & CL          & Purchase        & 0.7062       & +0.0038   \\ \bottomrule
  \end{tabular}
  }
  \end{table}
  \vspace{-1em}

\subsubsection{Effectiveness of C-SFT}
\label{sec:Effectiveness of C-SFT}
As presented in Section \ref{sec:C-SFT}, there are two important factors that influence C-SFT, i.e., the user interest signal and loss function.
Thus, we train different versions of C-SFT (see Table \ref{tab:Effectiveness Evaluation}) and evaluate the performance of these versions in CTR prediction tasks.
Specifically, for the user interest signal, we take the users' clicking  and users' purchase as the expression of user interest.
Furthermore, we also take an interest-irrelevant signal (item categories) with Cross Entropy loss as a comparison.
For the loss function, the contrastive loss (i.e., Equ \ref{equ:contrastive loss}) and Cross Entropy loss are compared.
The AUC results are reported in Table \ref{tab:Effectiveness Evaluation}.

\noindent \textbf{The impact of user interest signal. }
Compared with the performance of version v1, v2 and v3 achieves a better performance.
It demonstrates the importance of introducing the user interest signal.
Without such signals, MM features can hardly achieve a positive effect on CTR prediction.
Furthermore, v3 (or v5) obtains a better performance than v2 (or v4).
One of the possible reasons is that the purchase behavior is more strong and precise, resulting in a better expression of user interest.
It also shows that understanding user interest plays an important role in MIM.

\noindent \textbf{The impact of loss function. }
Compared with the baselines, both of the two loss functions perform better via bridging the semantic gap.
Furthermore, compared with the performance of version v2 (or v3), it shows that adopting contrastive loss (i.e., version v4 or v5) performs better, which indicates contrastive loss can better model the alignment between user interest and the target item, thereby further improving prediction performance.

\begin{table}[t] 
  \footnotesize
  \centering
  \caption{The results of ablation study.}
  \vspace{-1em}
  \label{tab:The results of ablation study.}
  \resizebox{0.65\columnwidth}{!}{
  \begin{tabular}{l|c}
  \toprule
  \textbf{Version}                      & \textbf{AUC}     \\ \midrule
  MIM                        & 0.7062  \\ \midrule
  \multicolumn{2}{l}{C-SFT stage}           \\  \midrule
  w/o TFN                       & 0.7058  \\
  w/o ST-NSG                    & 0.7056  \\
  w/o Multi-level InfoNCE       & 0.7052  \\
  w/o C-SFT                     & 0.7040  \\ \midrule
  \multicolumn{2}{l}{CiUBM} \\ \midrule
  w/o ID interest module        &  0.7057 \\
  w/o content interest module   &  0.7052 \\
  w/o fusion interest module    &  0.7054 \\
  \bottomrule
  \end{tabular}
  }
  \end{table}

\subsection{Ablation Study}
\label{sec:Ablation Study}
In this section, we conduct an ablation study to analyze the impact of different components.
The results are presented in Table \ref{tab:The results of ablation study.}.
Some observations are summarized as follows:  
1) For the C-SFT stage, it shows that C-SFT takes an important role and improves AUC from 0.7040 to 0.7062.
Besides, different components in C-SFT contribute to the final performance.
2) For the CiUBM, different interest modules contribute to the final improvement. 
It demonstrates that it is important to take both ID interest and content interest into consideration.
Note the analysis for the pre-training stage has been discussed in Section \ref{sec:Effectiveness of high-quality MM embeddings.}.

\subsection{Efficiency Evaluation}
\label{sec:Efficiency Evaluation}
In this section, we analyze the efficiency of the proposed paradigm, including time and GPU memory efficiency.
Detailedly, we develop two versions:
(1) MIM (w/o RC) refers to MIM without representation center, i.e., MM embedding is real-time inferred by FoM.
(2) MIM (E2E) refers to jointly training FoM and CiUBM in an E2E manner.
We report the GFLOPs (the number of giga floating-point operations) per sample to reflect the train and inference time cost and GPU memory usage in deepCTR \footnote{Since the pre-training stage can be pre-processed and C-SFT can be trained in parallel, the cost of these two stages can be ignored and is not included here.}.
The results are presented in Table \ref{tab:Efficiency Evaluation}.
We can find that 
(1) Compared with Base, both of MIM(w/o RC) and MIM(E2E)  need a large number of extra computations (773.55$\times$ $\sim$ 6318.12$\times$  higher than Base) and is GPU memory costly (8.23$\times$ $\sim$ 26.41$\times$  higher than Base).
Note the architecture of the MIM(w/o RC) and MIM(E2E) is the same, resulting in the same GFLOPs and GPU memory cost during inference.
(2) Compared with MIM(E2E), MIM(w/o RC) does not need to jointly train FoM and Base, which saves computation and GPU memory cost by the backward of FoM.
(3) For MIM, thanks to the decomposed paradigm and representation center, it has significantly reduced the cost introduced by FoM and only needs 0.59$\times$ and 0.387$\times$ extra computation in training and inference, respectively, and 0.01$\times$ and 0.81$\times$ extra GPU memory cost in training and inference, respectively.

Overall, such nice properties in terms of high performance, efficient GPU memory usage, and low time requirements of our proposed model are welcomed for web-scale applications.

  \begin{figure}[t]
  \centering
  \includegraphics[width=0.9\columnwidth]{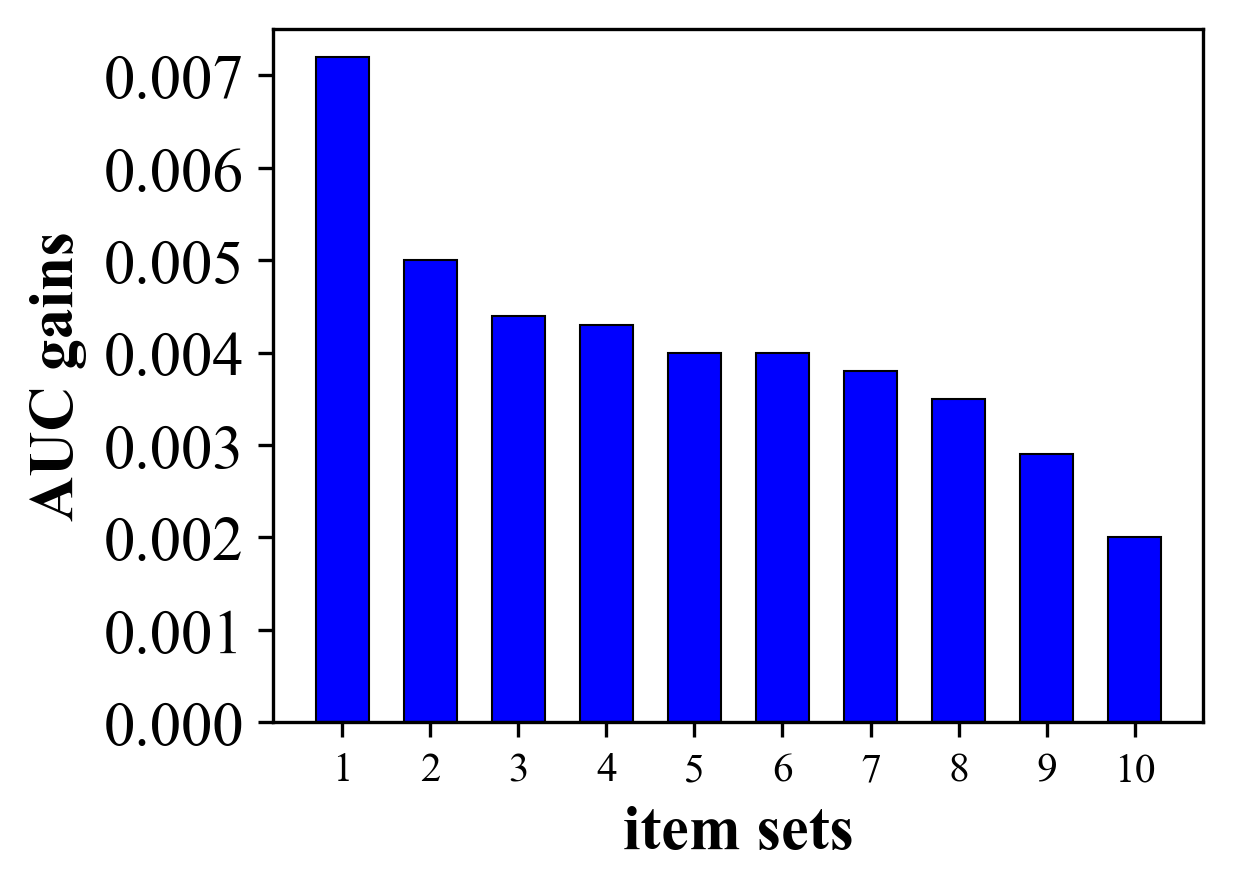}
  \caption{AUC gains by different item sets.}
  \label{fig:Auc gap by item sets}
  \end{figure}

  \begin{table}[t] \footnotesize
    \centering
    \caption{The results of efficiency evaluation. $\Delta=(Cost_A-Cost_B)/Cost_B$ is the relative difference with respect to Base.
    }
    \vspace{-1em}
    \centering
    \resizebox{\columnwidth}{!}{
        
    \label{tab:Efficiency Evaluation}
    \centering
    \begin{tabular}{l|cc|cc|cc|cc}
    \toprule
                        & \multicolumn{4}{c|}{\textbf{GFLOPs}}& \multicolumn{4}{c}{\textbf{GPU Memory (G)}} \\ \cline{2-9} 
    \multirow{3}{*}{{}} & \multicolumn{2}{c|}{\textbf{Train}}   & \multicolumn{2}{c|}{\textbf{Inference}}     & \multicolumn{2}{c|}{\textbf{Train}} & \multicolumn{2}{c}{\textbf{Inference}} \\ \cline{2-9}
                        & {Cost} & {$\Delta$} & {Cost} & {$\Delta$}& {Cost} & {$\Delta$}& {Cost} & {$\Delta$} \\ \hline
    Base              & {{0.114}}     & {{-}}           & {{0.31}}     & {-}          & {{1.1}}   & {{-}}        & {{0.68}} & {-}        \\ 
    Base+MIM(w/o RC)                 & 240.24& +2106.37& 240.11& +773.55 & {{10.15}} & {{+8.23}}  & {{9.85}} & {+13.49} \\
    Base+MIM(E2E)                 & 720.38& +6318.12& 240.11& +773.55 & {{30.15}} & {{+26.41}} & {{9.85}} & {+13.49} \\ 
    Base+MIM              & {{0.182}}     & {{+0.59}}       & {{0.43}}     & {+0.38.7}      & {{1.11}}  & {{+0.01}}    & {{0.81}} & {+0.19}   \\ \bottomrule
    \end{tabular}
    }
    \end{table}

\subsection{Generalization in Cold Start Scenarios}
\label{sec:Generalization in Cold Start Scenarios}
Here we analyze the performance of the cold start items to show the generalization of our proposed paradigm.
Detailedly, we divide different items into 10 sets (denoted as S1 to S10), where S1$\sim$S10 refers to the newest$\sim$oldest item sets.
Then we evaluate the improvement of different sets respectively.
The results are presented in Figure \ref{fig:Auc gap by item sets}.
It shows that MIM contributes more to the newest items (e.g., S1) and achieves more AUC gains. 
This demonstrates that MM features enable new items to better represent their features, leading to better performance.

\vspace{-1em}
\subsection{Evaluation of Industrial Applications}
\label{sec:Evaluation on Industrial Applications}
Here, we first develop MIM on a real-world industrial application, i.e., the sponsored search system in Taobao, and have continually achieved three releases of MIM in our industrial application in the past year.

\noindent \textbf{Online A/B Testing for CTR Prediction Tasks. } 
Compared with the online model, these three releases achieved 0.81pt AUC gains in total, and 0.26pt, 0.25pt, and 0.30pt AUC gains for each.
During the online A/B test, we observed 14.14\% CTR gains in total, and 5.1\%, 4.75\%, and 4.29\% CTR gains for each release.
It also achieves 4.12\% RPM (Revenue Per Mile) gain in total, and 1.5\%, 1.19\% and 1.43\% RPM gains for each release.

\noindent \textbf{Impact on Content-Sensitive Categories. }
We also report AUC, CTR, and RPM gains in different item categories in one of the releases (others have similar results).
Table \ref{tab:The improvement on different items.} shows more improvement can be archived on the content-sensitive categories (e.g., Clothes), where users more care about the styles of items and can be easily attracted by item images or titles.
On the contrary, lower improvement is obtained from content-insensitive categories (e.g., Cars).
It demonstrates user interest can be well modeled by MIM and improve the performance of CTR prediction.

\begin{table}[t]
  \footnotesize 
  \caption{The improvement on different item categories.}
  \vspace{-1em}
  \centering
  \resizebox{0.9\columnwidth}{!}{%
  \label{tab:The improvement on different items.}
  \begin{tabular}{l|c|c|c}
  \toprule
   \textbf{Category}                     & \textbf{AUC Gains} & \textbf{CTR Gains} & \textbf{RPM Gains} \\ \midrule
  \textbf{All}          & +0.0038            & +5\%               & +1\%               \\ \midrule
  \textbf{Home Decoration} & +0.0047            & +4.63\%            & +3.04\%            \\
  \textbf{Clothes}         & +0.0037            & +8.69\%            & +4.44\%            \\
  \textbf{Sports}          & +0.003             & +3.48\%            & +1.86\%            \\
  \textbf{Toys}            & +0.0024            & +4.44\%            & +1.89\%            \\
  \textbf{FMCG}            & +0.0028            & +3.98\%            & +2.74\%            \\
  \textbf{Pets}            & +0.0024            & +3.02\%            & +2.69\%            \\
  \textbf{Cars}            & +0.0028            & +2.54\%            & +0.02\%            \\ \bottomrule
  \end{tabular}%
  }
  \vspace{-1em}
  \end{table}
  
\noindent \textbf{Universality in Recommendation Tasks. }
Furthermore, we also try to apply MIM in a recommendation system, i.e., the display advertising system in Taobao, and achieve similar improvements.
Inspired by this method, the final deployed version achieved 0.4pt offline AUC gains, 3.5\% CTR gain and 1.5\% RPM gain online. From the success of different systems, we believe MIM is a universal and industrial welcomed multi-modal content interest modeling paradigm.

\section{Conclusion}
\label{sec:Conclusion}

In this paper, we propose MIM, a novel and universal multi-modal content interest modeling paradigm for industrial-scale applications. By introducing a decomposed training paradigm and a representation center, MIM effectively integrates multi-modal features into User Behavior Modeling (UBM), addressing the limitations of traditional ID-based methods and existing multi-modal approaches. The proposed method aligns user interests with content embeddings, significantly enhancing prediction performance while maintaining efficiency in large-scale real-world applications.
Extensive experiments validate the effectiveness and generalizability of MIM, showing consistent improvements across diverse baselines and datasets. These results demonstrate MIM's practical value and its potential to drive further innovation in multi-modal user modeling for industrial applications.

\bibliographystyle{ACM-Reference-Format}
\bibliography{mim}


\begin{thebibliography}{43}


\ifx \showCODEN    \undefined \def \showCODEN     #1{\unskip}     \fi
\ifx \showISBNx    \undefined \def \showISBNx     #1{\unskip}     \fi
\ifx \showISBNxiii \undefined \def \showISBNxiii  #1{\unskip}     \fi
\ifx \showISSN     \undefined \def \showISSN      #1{\unskip}     \fi
\ifx \showLCCN     \undefined \def \showLCCN      #1{\unskip}     \fi
\ifx \shownote     \undefined \def \shownote      #1{#1}          \fi
\ifx \showarticletitle \undefined \def \showarticletitle #1{#1}   \fi
\ifx \showURL      \undefined \def \showURL       {\relax}        \fi
\providecommand\bibfield[2]{#2}
\providecommand\bibinfo[2]{#2}
\providecommand\natexlab[1]{#1}
\providecommand\showeprint[2][]{arXiv:#2}

\bibitem[Alayrac et~al\mbox{.}(2022)]%
        {alayrac2022flamingo}
\bibfield{author}{\bibinfo{person}{Jean-Baptiste Alayrac}, \bibinfo{person}{Jeff Donahue}, \bibinfo{person}{Pauline Luc}, \bibinfo{person}{Antoine Miech}, \bibinfo{person}{Iain Barr}, \bibinfo{person}{Yana Hasson}, \bibinfo{person}{Karel Lenc}, \bibinfo{person}{Arthur Mensch}, \bibinfo{person}{Katherine Millican}, \bibinfo{person}{Malcolm Reynolds}, {et~al\mbox{.}}} \bibinfo{year}{2022}\natexlab{}.
\newblock \showarticletitle{Flamingo: a visual language model for few-shot learning}.
\newblock  (\bibinfo{year}{2022}).
\newblock


\bibitem[Bao et~al\mbox{.}(2021)]%
        {bao2021beit}
\bibfield{author}{\bibinfo{person}{Hangbo Bao}, \bibinfo{person}{Li Dong}, \bibinfo{person}{Songhao Piao}, {and} \bibinfo{person}{Furu Wei}.} \bibinfo{year}{2021}\natexlab{}.
\newblock \showarticletitle{Beit: Bert pre-training of image transformers}.
\newblock \bibinfo{journal}{\emph{arXiv preprint arXiv:2106.08254}} (\bibinfo{year}{2021}).
\newblock


\bibitem[Chang et~al\mbox{.}(2023)]%
        {chang2023twin}
\bibfield{author}{\bibinfo{person}{Jianxin Chang}, \bibinfo{person}{Chenbin Zhang}, \bibinfo{person}{Zhiyi Fu}, \bibinfo{person}{Xiaoxue Zang}, \bibinfo{person}{Lin Guan}, \bibinfo{person}{Jing Lu}, \bibinfo{person}{Yiqun Hui}, \bibinfo{person}{Dewei Leng}, \bibinfo{person}{Yanan Niu}, \bibinfo{person}{Yang Song}, {et~al\mbox{.}}} \bibinfo{year}{2023}\natexlab{}.
\newblock \showarticletitle{TWIN: TWo-stage interest network for lifelong user behavior modeling in CTR prediction at kuaishou}. In \bibinfo{booktitle}{\emph{Proceedings of the 29th ACM SIGKDD Conference on Knowledge Discovery and Data Mining}}. \bibinfo{pages}{3785--3794}.
\newblock


\bibitem[Chen et~al\mbox{.}(2020)]%
        {chen2020generative}
\bibfield{author}{\bibinfo{person}{Mark Chen}, \bibinfo{person}{Alec Radford}, \bibinfo{person}{Rewon Child}, \bibinfo{person}{Jeffrey Wu}, \bibinfo{person}{Heewoo Jun}, \bibinfo{person}{David Luan}, {and} \bibinfo{person}{Ilya Sutskever}.} \bibinfo{year}{2020}\natexlab{}.
\newblock \showarticletitle{Generative pretraining from pixels}. In \bibinfo{booktitle}{\emph{International conference on machine learning}}. PMLR, \bibinfo{pages}{1691--1703}.
\newblock


\bibitem[Chen et~al\mbox{.}(2019)]%
        {chen2019behavior}
\bibfield{author}{\bibinfo{person}{Qiwei Chen}, \bibinfo{person}{Huan Zhao}, \bibinfo{person}{Wei Li}, \bibinfo{person}{Pipei Huang}, {and} \bibinfo{person}{Wenwu Ou}.} \bibinfo{year}{2019}\natexlab{}.
\newblock \showarticletitle{Behavior sequence transformer for e-commerce recommendation in alibaba}. In \bibinfo{booktitle}{\emph{Proceedings of the 1st international workshop on deep learning practice for high-dimensional sparse data}}. \bibinfo{pages}{1--4}.
\newblock


\bibitem[Chiang et~al\mbox{.}(2023)]%
        {vicuna2023}
\bibfield{author}{\bibinfo{person}{Wei-Lin Chiang}, \bibinfo{person}{Zhuohan Li}, \bibinfo{person}{Zi Lin}, \bibinfo{person}{Ying Sheng}, \bibinfo{person}{Zhanghao Wu}, \bibinfo{person}{Hao Zhang}, \bibinfo{person}{Lianmin Zheng}, \bibinfo{person}{Siyuan Zhuang}, \bibinfo{person}{Yonghao Zhuang}, \bibinfo{person}{Joseph~E. Gonzalez}, \bibinfo{person}{Ion Stoica}, {and} \bibinfo{person}{Eric~P. Xing}.} \bibinfo{year}{2023}\natexlab{}.
\newblock \bibinfo{title}{Vicuna: An Open-Source Chatbot Impressing GPT-4 with 90\%* ChatGPT Quality}.
\newblock
\urldef\tempurl%
\url{https://lmsys.org/blog/2023-03-30-vicuna/}
\showURL{%
\tempurl}


\bibitem[Clark et~al\mbox{.}(2020)]%
        {DBLP:conf/iclr/ClarkLLM20}
\bibfield{author}{\bibinfo{person}{Kevin Clark}, \bibinfo{person}{Minh{-}Thang Luong}, \bibinfo{person}{Quoc~V. Le}, {and} \bibinfo{person}{Christopher~D. Manning}.} \bibinfo{year}{2020}\natexlab{}.
\newblock \showarticletitle{{ELECTRA:} Pre-training Text Encoders as Discriminators Rather Than Generators}. In \bibinfo{booktitle}{\emph{8th International Conference on Learning Representations, {ICLR} 2020, Addis Ababa, Ethiopia, April 26-30, 2020}}. \bibinfo{publisher}{OpenReview.net}.
\newblock
\urldef\tempurl%
\url{https://openreview.net/forum?id=r1xMH1BtvB}
\showURL{%
\tempurl}


\bibitem[Covington et~al\mbox{.}(2016)]%
        {covington2016Deep}
\bibfield{author}{\bibinfo{person}{Paul Covington}, \bibinfo{person}{Jay Adams}, {and} \bibinfo{person}{Emre Sargin}.} \bibinfo{year}{2016}\natexlab{}.
\newblock \showarticletitle{Deep {{Neural Networks}} for {{YouTube Recommendations}}}.
\newblock \bibinfo{journal}{\emph{Proceedings of the 10th ACM Conference on Recommender Systems}} (\bibinfo{date}{Sept.} \bibinfo{year}{2016}), \bibinfo{pages}{191--198}.
\newblock
\showISBNx{9781450340359}
\href{https://doi.org/10.1145/2959100.2959190}{doi:\nolinkurl{10.1145/2959100.2959190}}


\bibitem[Devlin et~al\mbox{.}(2018)]%
        {devlin2018bert}
\bibfield{author}{\bibinfo{person}{Jacob Devlin}, \bibinfo{person}{Ming-Wei Chang}, \bibinfo{person}{Kenton Lee}, {and} \bibinfo{person}{Kristina Toutanova}.} \bibinfo{year}{2018}\natexlab{}.
\newblock \showarticletitle{Bert: Pre-training of deep bidirectional transformers for language understanding}.
\newblock \bibinfo{journal}{\emph{arXiv preprint arXiv:1810.04805}} (\bibinfo{year}{2018}).
\newblock


\bibitem[Devlin et~al\mbox{.}(2019)]%
        {devlin2019BERT}
\bibfield{author}{\bibinfo{person}{Jacob Devlin}, \bibinfo{person}{Ming-Wei Chang}, \bibinfo{person}{Kenton Lee}, {and} \bibinfo{person}{Kristina Toutanova}.} \bibinfo{year}{2019}\natexlab{}.
\newblock \bibinfo{title}{{{BERT}}: {{Pre-training}} of {{Deep Bidirectional Transformers}} for {{Language Understanding}}}.
\newblock
\href{https://doi.org/10.48550/arXiv.1810.04805}{doi:\nolinkurl{10.48550/arXiv.1810.04805}}
\showeprint[arxiv]{1810.04805}~[cs]


\bibitem[Dosovitskiy et~al\mbox{.}(2020)]%
        {dosovitskiy2020Image}
\bibfield{author}{\bibinfo{person}{A. Dosovitskiy}, \bibinfo{person}{L. Beyer}, \bibinfo{person}{Alexander Kolesnikov}, \bibinfo{person}{Dirk Weissenborn}, \bibinfo{person}{Xiaohua Zhai}, \bibinfo{person}{Thomas Unterthiner}, \bibinfo{person}{Mostafa Dehghani}, \bibinfo{person}{Matthias Minderer}, \bibinfo{person}{G. Heigold}, \bibinfo{person}{S. Gelly}, \bibinfo{person}{Jakob Uszkoreit}, {and} \bibinfo{person}{N. Houlsby}.} \bibinfo{year}{2020}\natexlab{}.
\newblock \showarticletitle{An {{Image}} Is {{Worth}} 16x16 {{Words}}: {{Transformers}} for {{Image Recognition}} at {{Scale}}}.
\newblock \bibinfo{journal}{\emph{ArXiv}} (\bibinfo{date}{Oct.} \bibinfo{year}{2020}).
\newblock


\bibitem[Fang et~al\mbox{.}(2023)]%
        {fang2023eva02}
\bibfield{author}{\bibinfo{person}{Yuxin Fang}, \bibinfo{person}{Quan Sun}, \bibinfo{person}{Xinggang Wang}, \bibinfo{person}{Tiejun Huang}, \bibinfo{person}{Xinlong Wang}, {and} \bibinfo{person}{Yue Cao}.} \bibinfo{year}{2023}\natexlab{}.
\newblock \bibinfo{title}{EVA-02: A Visual Representation for Neon Genesis}.
\newblock
\showeprint[arxiv]{2303.11331}~[cs.CV]


\bibitem[Fawcett(2006)]%
        {fawcett2006introduction}
\bibfield{author}{\bibinfo{person}{Tom Fawcett}.} \bibinfo{year}{2006}\natexlab{}.
\newblock \showarticletitle{An introduction to ROC analysis}.
\newblock \bibinfo{journal}{\emph{Pattern recognition letters}} \bibinfo{volume}{27}, \bibinfo{number}{8} (\bibinfo{year}{2006}), \bibinfo{pages}{861--874}.
\newblock


\bibitem[He et~al\mbox{.}(2019)]%
        {he2019bag}
\bibfield{author}{\bibinfo{person}{Tong He}, \bibinfo{person}{Zhi Zhang}, \bibinfo{person}{Hang Zhang}, \bibinfo{person}{Zhongyue Zhang}, \bibinfo{person}{Junyuan Xie}, {and} \bibinfo{person}{Mu Li}.} \bibinfo{year}{2019}\natexlab{}.
\newblock \showarticletitle{Bag of tricks for image classification with convolutional neural networks}. In \bibinfo{booktitle}{\emph{Proceedings of the IEEE/CVF conference on computer vision and pattern recognition}}. \bibinfo{pages}{558--567}.
\newblock


\bibitem[Liu et~al\mbox{.}(2023)]%
        {liu2023Multimodal}
\bibfield{author}{\bibinfo{person}{Qidong Liu}, \bibinfo{person}{Jiaxi Hu}, \bibinfo{person}{Yutian Xiao}, \bibinfo{person}{Jingtong Gao}, {and} \bibinfo{person}{Xiang Zhao}.} \bibinfo{year}{2023}\natexlab{}.
\newblock \showarticletitle{Multimodal {{Recommender Systems}}: {{A Survey}}}.
\newblock \bibinfo{journal}{\emph{ArXiv}} (\bibinfo{year}{2023}).
\newblock


\bibitem[Liu et~al\mbox{.}(2021)]%
        {liu2021Swin}
\bibfield{author}{\bibinfo{person}{Ze Liu}, \bibinfo{person}{Yutong Lin}, \bibinfo{person}{Yue Cao}, \bibinfo{person}{Han Hu}, \bibinfo{person}{Yixuan Wei}, \bibinfo{person}{Zheng Zhang}, \bibinfo{person}{Stephen Lin}, {and} \bibinfo{person}{Baining Guo}.} \bibinfo{year}{2021}\natexlab{}.
\newblock \showarticletitle{Swin transformer: Hierarchical vision transformer using shifted windows}. In \bibinfo{booktitle}{\emph{Proceedings of the IEEE/CVF international conference on computer vision}}. \bibinfo{pages}{10012--10022}.
\newblock


\bibitem[Liu et~al\mbox{.}(2022)]%
        {liu2022ConvNeta}
\bibfield{author}{\bibinfo{person}{Zhuang Liu}, \bibinfo{person}{Hanzi Mao}, \bibinfo{person}{Chao-Yuan Wu}, \bibinfo{person}{Christoph Feichtenhofer}, \bibinfo{person}{Trevor Darrell}, {and} \bibinfo{person}{Saining Xie}.} \bibinfo{year}{2022}\natexlab{}.
\newblock \bibinfo{title}{A {{ConvNet}} for the 2020s}.
\newblock
\href{https://doi.org/10.48550/arXiv.2201.03545}{doi:\nolinkurl{10.48550/arXiv.2201.03545}}
\showeprint[arxiv]{2201.03545}~[cs]


\bibitem[Long et~al\mbox{.}(2024)]%
        {DBLP:journals/mta/LongDL24}
\bibfield{author}{\bibinfo{person}{Xianzhong Long}, \bibinfo{person}{Han Du}, {and} \bibinfo{person}{Yun Li}.} \bibinfo{year}{2024}\natexlab{}.
\newblock \showarticletitle{Two momentum contrast in triplet for unsupervised visual representation learning}.
\newblock \bibinfo{journal}{\emph{Multim. Tools Appl.}} \bibinfo{volume}{83}, \bibinfo{number}{4} (\bibinfo{year}{2024}), \bibinfo{pages}{10467--10480}.
\newblock
\href{https://doi.org/10.1007/S11042-023-15998-3}{doi:\nolinkurl{10.1007/S11042-023-15998-3}}


\bibitem[Lu et~al\mbox{.}(2019)]%
        {lu2019vilbert}
\bibfield{author}{\bibinfo{person}{Jiasen Lu}, \bibinfo{person}{Dhruv Batra}, \bibinfo{person}{Devi Parikh}, {and} \bibinfo{person}{Stefan Lee}.} \bibinfo{year}{2019}\natexlab{}.
\newblock \showarticletitle{Vilbert: Pretraining task-agnostic visiolinguistic representations for vision-and-language tasks}.
\newblock  (\bibinfo{year}{2019}).
\newblock


\bibitem[Mikolov et~al\mbox{.}(2013)]%
        {DBLP:conf/nips/MikolovSCCD13}
\bibfield{author}{\bibinfo{person}{Tom{\'{a}}s Mikolov}, \bibinfo{person}{Ilya Sutskever}, \bibinfo{person}{Kai Chen}, \bibinfo{person}{Gregory~S. Corrado}, {and} \bibinfo{person}{Jeffrey Dean}.} \bibinfo{year}{2013}\natexlab{}.
\newblock \showarticletitle{Distributed Representations of Words and Phrases and their Compositionality}. In \bibinfo{booktitle}{\emph{Advances in Neural Information Processing Systems 26: 27th Annual Conference on Neural Information Processing Systems 2013. Proceedings of a meeting held December 5-8, 2013, Lake Tahoe, Nevada, United States}}, \bibfield{editor}{\bibinfo{person}{Christopher J.~C. Burges}, \bibinfo{person}{L{\'{e}}on Bottou}, \bibinfo{person}{Zoubin Ghahramani}, {and} \bibinfo{person}{Kilian~Q. Weinberger}} (Eds.). \bibinfo{pages}{3111--3119}.
\newblock
\urldef\tempurl%
\url{https://proceedings.neurips.cc/paper/2013/hash/9aa42b31882ec039965f3c4923ce901b-Abstract.html}
\showURL{%
\tempurl}


\bibitem[Mo et~al\mbox{.}(2015)]%
        {mo2015image}
\bibfield{author}{\bibinfo{person}{Kaixiang Mo}, \bibinfo{person}{Bo Liu}, \bibinfo{person}{Lei Xiao}, \bibinfo{person}{Yong Li}, {and} \bibinfo{person}{Jie Jiang}.} \bibinfo{year}{2015}\natexlab{}.
\newblock \showarticletitle{Image feature learning for cold start problem in display advertising}. In \bibinfo{booktitle}{\emph{Twenty-Fourth International Joint Conference on Artificial Intelligence}}.
\newblock


\bibitem[OpenAI et~al\mbox{.}(2024)]%
        {openai2024gpt4technicalreport}
\bibfield{author}{\bibinfo{person}{OpenAI}, \bibinfo{person}{Josh Achiam}, \bibinfo{person}{Steven Adler}, \bibinfo{person}{Sandhini Agarwal}, {and} \bibinfo{person}{Lama Ahmad...}} \bibinfo{year}{2024}\natexlab{}.
\newblock \bibinfo{title}{GPT-4 Technical Report}.
\newblock
\showeprint[arxiv]{2303.08774}~[cs.CL]
\urldef\tempurl%
\url{https://arxiv.org/abs/2303.08774}
\showURL{%
\tempurl}


\bibitem[Pi et~al\mbox{.}(2020)]%
        {pi2020search}
\bibfield{author}{\bibinfo{person}{Qi Pi}, \bibinfo{person}{Guorui Zhou}, \bibinfo{person}{Yujing Zhang}, \bibinfo{person}{Zhe Wang}, \bibinfo{person}{Lejian Ren}, \bibinfo{person}{Ying Fan}, \bibinfo{person}{Xiaoqiang Zhu}, {and} \bibinfo{person}{Kun Gai}.} \bibinfo{year}{2020}\natexlab{}.
\newblock \showarticletitle{Search-based user interest modeling with lifelong sequential behavior data for click-through rate prediction}. In \bibinfo{booktitle}{\emph{Proceedings of the 29th ACM International Conference on Information \& Knowledge Management}}. \bibinfo{pages}{2685--2692}.
\newblock


\bibitem[Qiu et~al\mbox{.}(2020)]%
        {DBLP:conf/kdd/QiuCDZYDWT20}
\bibfield{author}{\bibinfo{person}{Jiezhong Qiu}, \bibinfo{person}{Qibin Chen}, \bibinfo{person}{Yuxiao Dong}, \bibinfo{person}{Jing Zhang}, \bibinfo{person}{Hongxia Yang}, \bibinfo{person}{Ming Ding}, \bibinfo{person}{Kuansan Wang}, {and} \bibinfo{person}{Jie Tang}.} \bibinfo{year}{2020}\natexlab{}.
\newblock \showarticletitle{{GCC:} Graph Contrastive Coding for Graph Neural Network Pre-Training}. In \bibinfo{booktitle}{\emph{{KDD} '20: The 26th {ACM} {SIGKDD} Conference on Knowledge Discovery and Data Mining, Virtual Event, CA, USA, August 23-27, 2020}}, \bibfield{editor}{\bibinfo{person}{Rajesh Gupta}, \bibinfo{person}{Yan Liu}, \bibinfo{person}{Jiliang Tang}, {and} \bibinfo{person}{B.~Aditya Prakash}} (Eds.). \bibinfo{publisher}{{ACM}}, \bibinfo{pages}{1150--1160}.
\newblock
\href{https://doi.org/10.1145/3394486.3403168}{doi:\nolinkurl{10.1145/3394486.3403168}}


\bibitem[Radford et~al\mbox{.}(2021)]%
        {radford2021learning}
\bibfield{author}{\bibinfo{person}{Alec Radford}, \bibinfo{person}{Jong~Wook Kim}, \bibinfo{person}{Chris Hallacy}, \bibinfo{person}{Aditya Ramesh}, \bibinfo{person}{Gabriel Goh}, \bibinfo{person}{Sandhini Agarwal}, \bibinfo{person}{Girish Sastry}, \bibinfo{person}{Amanda Askell}, \bibinfo{person}{Pamela Mishkin}, \bibinfo{person}{Jack Clark}, {et~al\mbox{.}}} \bibinfo{year}{2021}\natexlab{}.
\newblock \showarticletitle{Learning transferable visual models from natural language supervision}. In \bibinfo{booktitle}{\emph{International conference on machine learning}}. PMLR, \bibinfo{pages}{8748--8763}.
\newblock


\bibitem[Radford et~al\mbox{.}(2019)]%
        {radford2019language}
\bibfield{author}{\bibinfo{person}{Alec Radford}, \bibinfo{person}{Jeffrey Wu}, \bibinfo{person}{Rewon Child}, \bibinfo{person}{David Luan}, \bibinfo{person}{Dario Amodei}, \bibinfo{person}{Ilya Sutskever}, {et~al\mbox{.}}} \bibinfo{year}{2019}\natexlab{}.
\newblock \showarticletitle{Language models are unsupervised multitask learners}.
\newblock \bibinfo{journal}{\emph{OpenAI blog}} \bibinfo{volume}{1}, \bibinfo{number}{8} (\bibinfo{year}{2019}), \bibinfo{pages}{9}.
\newblock


\bibitem[Tan and Bansal(2019)]%
        {tan2019lxmert}
\bibfield{author}{\bibinfo{person}{Hao Tan} {and} \bibinfo{person}{Mohit Bansal}.} \bibinfo{year}{2019}\natexlab{}.
\newblock \showarticletitle{Lxmert: Learning cross-modality encoder representations from transformers}.
\newblock \bibinfo{journal}{\emph{arXiv preprint arXiv:1908.07490}} (\bibinfo{year}{2019}).
\newblock


\bibitem[Touvron et~al\mbox{.}(2023)]%
        {touvron2023llamaopenefficientfoundation}
\bibfield{author}{\bibinfo{person}{Hugo Touvron}, \bibinfo{person}{Thibaut Lavril}, \bibinfo{person}{Gautier Izacard}, \bibinfo{person}{Xavier Martinet}, \bibinfo{person}{Marie-Anne Lachaux}, \bibinfo{person}{Timothée Lacroix}, \bibinfo{person}{Baptiste Rozière}, \bibinfo{person}{Naman Goyal}, \bibinfo{person}{Eric Hambro}, \bibinfo{person}{Faisal Azhar}, \bibinfo{person}{Aurelien Rodriguez}, \bibinfo{person}{Armand Joulin}, \bibinfo{person}{Edouard Grave}, {and} \bibinfo{person}{Guillaume Lample}.} \bibinfo{year}{2023}\natexlab{}.
\newblock \bibinfo{title}{LLaMA: Open and Efficient Foundation Language Models}.
\newblock
\showeprint[arxiv]{2302.13971}~[cs.CL]
\urldef\tempurl%
\url{https://arxiv.org/abs/2302.13971}
\showURL{%
\tempurl}


\bibitem[Wang et~al\mbox{.}(2023)]%
        {wang2023missrec}
\bibfield{author}{\bibinfo{person}{Jinpeng Wang}, \bibinfo{person}{Ziyun Zeng}, \bibinfo{person}{Yunxiao Wang}, \bibinfo{person}{Yuting Wang}, \bibinfo{person}{Xingyu Lu}, \bibinfo{person}{Tianxiang Li}, \bibinfo{person}{Jun Yuan}, \bibinfo{person}{Rui Zhang}, \bibinfo{person}{Hai-Tao Zheng}, {and} \bibinfo{person}{Shu-Tao Xia}.} \bibinfo{year}{2023}\natexlab{}.
\newblock \showarticletitle{MISSRec: Pre-training and Transferring Multi-modal Interest-aware Sequence Representation for Recommendation}. In \bibinfo{booktitle}{\emph{Proceedings of the 31st ACM International Conference on Multimedia}}. \bibinfo{pages}{6548--6557}.
\newblock


\bibitem[Wang et~al\mbox{.}(2022)]%
        {wang2022image}
\bibfield{author}{\bibinfo{person}{Wenhui Wang}, \bibinfo{person}{Hangbo Bao}, \bibinfo{person}{Li Dong}, \bibinfo{person}{Johan Bjorck}, \bibinfo{person}{Zhiliang Peng}, \bibinfo{person}{Qiang Liu}, \bibinfo{person}{Kriti Aggarwal}, \bibinfo{person}{Owais~Khan Mohammed}, \bibinfo{person}{Saksham Singhal}, \bibinfo{person}{Subhojit Som}, {et~al\mbox{.}}} \bibinfo{year}{2022}\natexlab{}.
\newblock \showarticletitle{Image as a foreign language: Beit pretraining for all vision and vision-language tasks}.
\newblock \bibinfo{journal}{\emph{arXiv preprint arXiv:2208.10442}} (\bibinfo{year}{2022}).
\newblock


\bibitem[Woo et~al\mbox{.}(2023)]%
        {woo2023convnext}
\bibfield{author}{\bibinfo{person}{Sanghyun Woo}, \bibinfo{person}{Shoubhik Debnath}, \bibinfo{person}{Ronghang Hu}, \bibinfo{person}{Xinlei Chen}, \bibinfo{person}{Zhuang Liu}, \bibinfo{person}{In~So Kweon}, {and} \bibinfo{person}{Saining Xie}.} \bibinfo{year}{2023}\natexlab{}.
\newblock \bibinfo{title}{ConvNeXt V2: Co-designing and Scaling ConvNets with Masked Autoencoders}.
\newblock
\showeprint[arxiv]{2301.00808}~[cs.CV]


\bibitem[Wu et~al\mbox{.}(2021)]%
        {wu2021empowering}
\bibfield{author}{\bibinfo{person}{Chuhan Wu}, \bibinfo{person}{Fangzhao Wu}, \bibinfo{person}{Tao Qi}, {and} \bibinfo{person}{Yongfeng Huang}.} \bibinfo{year}{2021}\natexlab{}.
\newblock \showarticletitle{Empowering news recommendation with pre-trained language models}. In \bibinfo{booktitle}{\emph{Proceedings of the 44th International ACM SIGIR Conference on Research and Development in Information Retrieval}}. \bibinfo{pages}{1652--1656}.
\newblock


\bibitem[Wu et~al\mbox{.}(2018)]%
        {DBLP:conf/cvpr/WuXYL18}
\bibfield{author}{\bibinfo{person}{Zhirong Wu}, \bibinfo{person}{Yuanjun Xiong}, \bibinfo{person}{Stella~X. Yu}, {and} \bibinfo{person}{Dahua Lin}.} \bibinfo{year}{2018}\natexlab{}.
\newblock \showarticletitle{Unsupervised Feature Learning via Non-Parametric Instance Discrimination}. In \bibinfo{booktitle}{\emph{2018 {IEEE} Conference on Computer Vision and Pattern Recognition, {CVPR} 2018, Salt Lake City, UT, USA, June 18-22, 2018}}. \bibinfo{publisher}{Computer Vision Foundation / {IEEE} Computer Society}, \bibinfo{pages}{3733--3742}.
\newblock
\href{https://doi.org/10.1109/CVPR.2018.00393}{doi:\nolinkurl{10.1109/CVPR.2018.00393}}


\bibitem[Xiao et~al\mbox{.}(2023)]%
        {bge_embedding}
\bibfield{author}{\bibinfo{person}{Shitao Xiao}, \bibinfo{person}{Zheng Liu}, \bibinfo{person}{Peitian Zhang}, {and} \bibinfo{person}{Niklas Muennighoff}.} \bibinfo{year}{2023}\natexlab{}.
\newblock \bibinfo{title}{C-Pack: Packaged Resources To Advance General Chinese Embedding}.
\newblock
\showeprint[arxiv]{2309.07597}~[cs.CL]


\bibitem[Yan et~al\mbox{.}(2022)]%
        {yan2022apg}
\bibfield{author}{\bibinfo{person}{Bencheng Yan}, \bibinfo{person}{Pengjie Wang}, \bibinfo{person}{Kai Zhang}, \bibinfo{person}{Feng Li}, \bibinfo{person}{Hongbo Deng}, \bibinfo{person}{Jian Xu}, {and} \bibinfo{person}{Bo Zheng}.} \bibinfo{year}{2022}\natexlab{}.
\newblock \showarticletitle{Apg: Adaptive parameter generation network for click-through rate prediction}.
\newblock \bibinfo{journal}{\emph{Advances in Neural Information Processing Systems}}  \bibinfo{volume}{35} (\bibinfo{year}{2022}), \bibinfo{pages}{24740--24752}.
\newblock


\bibitem[Yang et~al\mbox{.}(2019)]%
        {yang2019xlnet}
\bibfield{author}{\bibinfo{person}{Zhilin Yang}, \bibinfo{person}{Zihang Dai}, \bibinfo{person}{Yiming Yang}, \bibinfo{person}{Jaime Carbonell}, \bibinfo{person}{Russ~R Salakhutdinov}, {and} \bibinfo{person}{Quoc~V Le}.} \bibinfo{year}{2019}\natexlab{}.
\newblock \showarticletitle{Xlnet: Generalized autoregressive pretraining for language understanding}.
\newblock \bibinfo{journal}{\emph{Advances in neural information processing systems}}  \bibinfo{volume}{32} (\bibinfo{year}{2019}).
\newblock


\bibitem[Yuan et~al\mbox{.}(2020)]%
        {yuan2020ParameterEfficient}
\bibfield{author}{\bibinfo{person}{Fajie Yuan}, \bibinfo{person}{Xiangnan He}, \bibinfo{person}{Alexandros Karatzoglou}, {and} \bibinfo{person}{Liguang Zhang}.} \bibinfo{year}{2020}\natexlab{}.
\newblock \bibinfo{title}{Parameter-{{Efficient Transfer}} from {{Sequential Behaviors}} for {{User Modeling}} and {{Recommendation}}}.
\newblock
\showeprint[arxiv]{2001.04253}~[cs]


\bibitem[Yuan et~al\mbox{.}(2021)]%
        {yuan2021One}
\bibfield{author}{\bibinfo{person}{Fajie Yuan}, \bibinfo{person}{Guoxiao Zhang}, \bibinfo{person}{Alexandros Karatzoglou}, \bibinfo{person}{Joemon Jose}, \bibinfo{person}{Beibei Kong}, {and} \bibinfo{person}{Yudong Li}.} \bibinfo{year}{2021}\natexlab{}.
\newblock \showarticletitle{One {{Person}}, {{One Model}}, {{One World}}: {{Learning Continual User Representation}} without {{Forgetting}}}.
\newblock \bibinfo{journal}{\emph{Proceedings of the 44th International ACM SIGIR Conference on Research and Development in Information Retrieval}} (\bibinfo{date}{July} \bibinfo{year}{2021}), \bibinfo{pages}{696--705}.
\newblock
\showISBNx{9781450380379}
\href{https://doi.org/10.1145/3404835.3462884}{doi:\nolinkurl{10.1145/3404835.3462884}}


\bibitem[Yuan et~al\mbox{.}(2023)]%
        {yuan2023Where}
\bibfield{author}{\bibinfo{person}{Zheng Yuan}, \bibinfo{person}{Fajie Yuan}, \bibinfo{person}{Yu Song}, \bibinfo{person}{Youhua Li}, \bibinfo{person}{Junchen Fu}, \bibinfo{person}{Fei Yang}, \bibinfo{person}{Yunzhu Pan}, {and} \bibinfo{person}{Yongxin Ni}.} \bibinfo{year}{2023}\natexlab{}.
\newblock \showarticletitle{Where to {{Go Next}} for {{Recommender Systems}}? {{ID-}} vs. {{Modality-based Recommender Models Revisited}}}.
\newblock \bibinfo{journal}{\emph{Proceedings of the 46th International ACM SIGIR Conference on Research and Development in Information Retrieval}} (\bibinfo{date}{July} \bibinfo{year}{2023}), \bibinfo{pages}{2639--2649}.
\newblock
\showISBNx{9781450394086}
\href{https://doi.org/10.1145/3539618.3591932}{doi:\nolinkurl{10.1145/3539618.3591932}}


\bibitem[Zadeh et~al\mbox{.}(2017)]%
        {zadeh2017tensor}
\bibfield{author}{\bibinfo{person}{Amir Zadeh}, \bibinfo{person}{Minghai Chen}, \bibinfo{person}{Soujanya Poria}, \bibinfo{person}{Erik Cambria}, {and} \bibinfo{person}{Louis-Philippe Morency}.} \bibinfo{year}{2017}\natexlab{}.
\newblock \showarticletitle{Tensor Fusion Network for Multimodal Sentiment Analysis}. In \bibinfo{booktitle}{\emph{Proceedings of the 2017 Conference on Empirical Methods in Natural Language Processing}}. \bibinfo{pages}{1103--1114}.
\newblock


\bibitem[Zhang et~al\mbox{.}(2021)]%
        {zhang2021deep}
\bibfield{author}{\bibinfo{person}{Weinan Zhang}, \bibinfo{person}{Jiarui Qin}, \bibinfo{person}{Wei Guo}, \bibinfo{person}{Ruiming Tang}, {and} \bibinfo{person}{Xiuqiang He}.} \bibinfo{year}{2021}\natexlab{}.
\newblock \showarticletitle{Deep learning for click-through rate estimation}.
\newblock \bibinfo{journal}{\emph{arXiv preprint arXiv:2104.10584}} (\bibinfo{year}{2021}).
\newblock


\bibitem[Zhou et~al\mbox{.}(2019)]%
        {zhou2019deep}
\bibfield{author}{\bibinfo{person}{Guorui Zhou}, \bibinfo{person}{Na Mou}, \bibinfo{person}{Ying Fan}, \bibinfo{person}{Qi Pi}, \bibinfo{person}{Weijie Bian}, \bibinfo{person}{Chang Zhou}, \bibinfo{person}{Xiaoqiang Zhu}, {and} \bibinfo{person}{Kun Gai}.} \bibinfo{year}{2019}\natexlab{}.
\newblock \showarticletitle{Deep interest evolution network for click-through rate prediction}. In \bibinfo{booktitle}{\emph{Proceedings of the AAAI conference on artificial intelligence}}, Vol.~\bibinfo{volume}{33}. \bibinfo{pages}{5941--5948}.
\newblock


\bibitem[Zhou et~al\mbox{.}(2018)]%
        {zhou2018deep}
\bibfield{author}{\bibinfo{person}{Guorui Zhou}, \bibinfo{person}{Xiaoqiang Zhu}, \bibinfo{person}{Chenru Song}, \bibinfo{person}{Ying Fan}, \bibinfo{person}{Han Zhu}, \bibinfo{person}{Xiao Ma}, \bibinfo{person}{Yanghui Yan}, \bibinfo{person}{Junqi Jin}, \bibinfo{person}{Han Li}, {and} \bibinfo{person}{Kun Gai}.} \bibinfo{year}{2018}\natexlab{}.
\newblock \showarticletitle{Deep interest network for click-through rate prediction}. In \bibinfo{booktitle}{\emph{Proceedings of the 24th ACM SIGKDD international conference on knowledge discovery \& data mining}}. \bibinfo{pages}{1059--1068}.
\newblock


\end{thebibliography}










\end{document}